\documentclass[aps,prd,groupedaddress,showpacs,showkeys]{revtex4}
\usepackage{amssymb}
\usepackage{epsfig}

\newcommand{\eq}{\begin{eqnarray}}
\newcommand{\en}{\end{eqnarray}}
\newcommand{\bea}{\begin{eqnarray}}
\newcommand{\eea}{\end{eqnarray}}

\newcommand{\ra}{\rangle}
\newcommand{\la}{\langle}
\newcommand{\mnz}{\stackrel{\!\!\!\!\!\circ}{m_N}}

\begin{document}

\title{Light baryon magnetic moments
       and $N\to \Delta \gamma$ transition \\
       in a Lorentz covariant chiral quark approach}

\noindent 
\author{Amand Faessler$^{1}$, 
        Thomas Gutsche$^{1}$,
        Barry R. Holstein$^{2}$, 
        Valery E. Lyubovitskij$^1$\footnote{On leave of absence  
        from Department of Physics, Tomsk State University, 
        634050 Tomsk, Russia}, 
        Diana Nicmorus$^1$\footnote{On leave of absence 
        from Institute of Space Sciences, P.O. Box MG-23, 
        Bucharest-Magurele 76900 Romania}, 
        Kem Pumsa-ard$^{1}$
\vspace*{1.2\baselineskip}}

\affiliation{$^1$ Institut f\"ur Theoretische Physik,
Universit\"at T\"ubingen,\\
Auf der Morgenstelle 14, D-72076 T\"ubingen, Germany
\vspace*{1.2\baselineskip} \\
$^2$ Department of Physics--LGRT, University of
Massachusetts, Amherst, MA 01003 USA\\}

\date{\today}

\begin{abstract}

We calculate magnetic moments of light baryons and
$N \to \Delta \gamma$ transition characteristics using
a manifestly Lorentz covariant chiral quark approach for the
study of baryons as bound states of constituent quarks dressed
by a cloud of pseudoscalar mesons.

\end{abstract}



\maketitle

\newpage

\section{Introduction}

The study of the magnetic moments of light baryons and of the
$N \to \Delta \gamma$ transition represents an old and important
problem in hadron physics. Many theoretical approaches---lattice
QCD, QCD sum rules, Chiral Perturbation Theory (ChPT),
various quark and soliton methods, techniques based on the
solution of Bethe-Salpeter and Faddeev field equations,
etc. -- have been applied in order to calculate these quantities.

It should be stressed that analysis of the $N \to \Delta \gamma$
transition is of particular interest because it allows one to
probe the structure of both the nucleon and $\Delta(1232)$-isobar
and can help to shed light on their possible deformation. This
reaction represents a crucial test for the various theoretical
approaches. For example, naive quark models based on SU(6)
symmetry, which model the nucleon and its first resonance as a
spherically symmetric 3q-configurations, fail to correctly
describe the electric $G_{E2}$ and Coulomb $G_{C2}$ quadrupole
form factors, which vanish in such models in contradistinction
with experiment.

In Refs.~\cite{Bjorken:1966ij,Jones:1972ky,Devenish:1975jd}
a model-independent analysis of the $N \to \Delta \gamma$ transition
amplitude has been performed. Based on gauge and Lorentz
covariance it was shown that the corresponding vertex function can
be expressed in terms of three linearly independent form factors.
All aspects of the reaction, such as helicity or multipole
amplitudes, are expressible in terms of these form factors.

A comprehensive review of the role of nucleon resonances in nuclear
structure has been presented in Ref.~\cite{Weber:1978dh}.
A didactic introduction to the $N-\Delta$ transition involving
the main theoretical ideas and predictions of the constituent
quark model (CQM) and its applications to the electromagnetic
properties of nucleons and nuclei is given in
Ref.~\cite{Giannini:1990pc}. This paper reviews the Isgur-Karl
model and presents basic formulae for calculation of the baryon
spectrum. Quarks are fundamental carriers of the baryon charge
and coupling of the photon is introduced at the quark level. The
model is used in the evaluation of the electromagnetic properties
in terms of nucleon form factors, the $\Delta$ electromagnetic
form factors, and excitations of the nucleon resonances. Further
improvements are proposed: inclusion of relativistic effects,
introduction of pion degrees of freedom, etc.

An effective Lagrangian incorporating chiral symmetry has been
utilized in~\cite{Davidson:1991xz}. This Lagrangian includes at
tree level the pseudo-vector Born terms, leading $t-$channel
vector-meson exchanges, as well as $s-$ and $u-$channel
$\Delta$-isobar exchanges. The magnetic dipole ($M1$) and electric
quadrupole ($E2$) amplitudes are expressed in terms of two
independent gauge couplings at the $\gamma N \Delta $ vertex.
The investigation of pion photoproduction from threshold through
the $\Delta(1232)$ resonance region is accomplished using various
unitarization methods, such that the errors obtained for both $E2$
and $M1$ multipoles reflect theoretical uncertainties as well as
model dependence.

Ref.~\cite{Hemmert:1994ky} analyzed the vector and axial form factors
of the $NN$ and $N\Delta$ systems as well as the $\pi NN$ and
$\pi N \Delta$ coupling constants (calculated defining two effective
Lagrangians for the $\pi NN$ and $\pi N \Delta$ interactions) within
a constituent quark model. The main conclusion is that while the
Goldberger-Treiman relation remains valid, the experimental
couplings are found to be larger by 30\% or so than those
predicted by the model. Also, the use of a constituent quark
model provides significant mass-dependent corrections to the naive
predictions of SU(6) symmetry.

Complex form factors were calculated to order $O(\epsilon^{3})$ in
the "small scale expansion" formalism (inclusion of the $\Delta$
degrees of freedom consistent with chiral symmetry), within the
framework of chiral effective theory~\cite{Gellas:1998wx}.
It is shown that the low-$q^2$ dependence of the three transition
multipoles -- $M1(q^2)$, $E2(q^2)$ and $C2(q^2)$ -- is governed by
the $\pi N$ and $\pi \Delta$ loop effects. The effective chiral
lagrangian incorporates both the spontaneous and explicit breaking
of chiral invariance.  The way in which unknown low energy
constants affect the ratios EMR$(q^2)=E2(q^2)/M1(q^2)$ and
CMR$(q^2)=C2(q^2)/M1(q^2)$ is elucidated, and estimated values
for the three individual couplings are obtained.

In Ref.~\cite{Buchmann:2004ia} it was demonstrated that the
$C2/M1$ ratio is related to the neutron elastic form factor ratio
$G_{C}^{n}/G_{M}^{n}$ not only at zero momentum transfer, but also
for the entire range of momentum transfer where data is available.
Relations are presented between the charge quadrupole transition
form factor and elastic nucleon charge form factor on one side and
the magnetic dipole transition form factor and elastic
neutron magnetic form factor on the other. For example,
at zero momentum transfer, the transition quadrupole moment and
the neutron charge radius are related, leading the authors to the
conclusion that the phenomena of deviation from the nucleon's spherical
symmetry has its origin in a nonspherical cloud of quark-antiquark pairs
in the nucleon. Performing an extrapolation of the $C2/M1$ result to
$Q^2 \rightarrow  \infty $ the ratio asymptotically approaches
a small negative constant in qualitative agreement with
perturbative QCD (pQCD).

Ref.~\cite{Pascalutsa:2005ts} studied the chiral behavior
($M_{\pi}$ dependence) of the $\gamma N \Delta$ EMR and CMR ratios
using a relativistic effective chiral Lagrangian involving pion
and nucleon fields supplemented by relativistic $\Delta$-isobar
fields. The calculation of observables in the pion
electroproduction amplitude was performed to next-to-leading order
(NLO) in the $\delta$-expansion. The parameters entering the
calculation of the various cross sections are the couplings
$g_{M}, g_{E}$ and $g_{C}$ characterizing the individual $M1, E2$
and $C2$ transitions.

In Ref.~\cite{Braun:2005be} a theoretical framework using the
light-cone sum rule approach has been suggested for the
calculation of the $\gamma^*N \rightarrow \Delta$ transition.
All three possibilities for virtual photon polarization were
allowed, so the transition is described by three independent form
factors.  Since predictions are close to data in the region above
$Q^2 \sim 2$ GeV$^2$, the main conclusion on the result for the
magnetic form factor is that the ''soft'' contribution is dominant
at the experimentally accessible momentum transfers.

There are a number of interesting problems which we
address in the present paper:

\begin{itemize}

\item [i)] if one believes that both valence and sea-quark effects
are important in the description of the electromagnetic properties
of light baryons, then how large is the contribution of the
meson-cloud;

\item [ii)] what is the physics required to correctly predict the $M1$
amplitude for the $N\to\Delta$ transition, which is considerably
underestimated in constituent quark models;

\item [iii)] what input is needed in order to explain the experimental
data for $E2/M1$ and $C2/M1$.

\end{itemize}

To possibly answer the above questions we use a Lorentz covariant
chiral quark model recently developed in Ref.~\cite{Faessler:2005gd}.
The approach is based on a non-linear chirally symmetric
Lagrangian, which involves constituent quarks and the chiral
(pseudoscalar meson) fields as the effective degrees of
freedom. In a first step, this Lagrangian can be used to perform
a dressing of the constituent quarks by a cloud of light
pseudoscalar mesons and other heavy states using the calculational
technique of infrared dimensional regularization (IDR) of loop
diagrams. Then within a proper chiral expansion, we calculate the
dressed transition operators which are relevant for the
interaction of the quarks with external fields in the presence of
a virtual meson cloud. In a following step, these dressed
operators are used to calculate baryon matrix elements. Note,
that a simpler and more phenomenological quark
model which was based on the similar ideas of the dressing of the
constituent quarks by a meson cloud has been developed in
Refs.~\cite{PCQM}.

In the manuscript we proceed as follows. First, in Section II, we
discuss basic notions of our approach. We derive the chiral
Lagrangian motivated by baryon
ChPT~\cite{Becher:1999he}-\cite{Fuchs:2003kq}, and
formulate it in terms of quark and mesonic degrees of freedom.
Next, we use this Lagrangian to perform
a dressing of the constituent quarks by a cloud of light pseudoscalar
mesons and by other heavy states, using the calculational
technique developed in Ref.~\cite{Becher:1999he}.
We derive dressed transition operators within a proper chiral
expansion, which are in turn relevant for the interaction of
quarks with external fields in the presence of a virtual meson
cloud. Then we discuss the calculation of matrix elements of dressed
quark operators between baryons states using a specific
constituent quark model~\cite{Ivanov:1996pz}-\cite{Faessler:2006ft}
based on a specific hadronization ansatz of quarks in baryons.
In Section III, we apply our approach to the study of magnetic
moments of light baryons (nucleons and hyperons) and to the
properties of the $N\to\Delta\gamma$ transition.
In Section IV we present a short summary of our results.

\section{Approach}

\subsection{Chiral Lagrangian}

\noindent

The chiral quark Lagrangian ${\cal L}_{qU}$ (up to order $p^4$),
which dynamically generates the dressing of the constituent quarks
by  mesonic degrees of freedom, consists of two primary pieces
${\cal L}_{q}$ and~${\cal L}_{U}$:
\eq\label{L_qU}
{\cal L}_{qU} \, = \, {\cal L}_{q} + {\cal L}_{U}\,,
\hspace*{.5cm}
{\cal L}_q \, = \, {\cal L}^{(1)}_q + {\cal L}^{(2)}_q +
{\cal L}^{(3)}_q + {\cal L}^{(4)}_q + \cdots\,,
\hspace*{.5cm}
{\cal L}_{U} \, = \, {\cal L}_{U}^{(2)} + \cdots\,.
\en
The superscript $(i)$ attached to ${\cal L}^{(i)}_{q(U)}$
denotes the low energy dimension of the Lagrangian:
\eq\label{L_exp}
{\cal L}_{U}^{(2)} &=&\frac{F^2}{4} \la{u_\mu u^\mu + \chi_+}\ra\,,
\hspace*{.5cm}
{\cal L}^{(1)}_q \, = \,  \bar q \left[ i \, \slash\!\!\!\! D - m
+ \frac{1}{2} \, g \, \slash\!\!\! u \, \gamma^5 \right] q\,,
\nonumber\\[2mm]
{\cal L}^{(2)}_q & = & - \frac{c_2}{4m^2} \, \la{u_\mu u_\nu}\ra \,
(\bar q \, D^\mu \, D^\nu \, q \, + \, {\, \rm h.c. \,}) \, + \,
\frac{c_4}{4}\, \bar q\, i\, \sigma^{\mu\nu}\, [u_\mu, u_\nu]\, q
\, + \, \frac{c_6}{8m}\,\bar q\,\sigma^{\mu\nu}\, F_{\mu\nu}^+ \, q\,
\, + \, \cdots , \\[2mm]
{\cal L}^{(3)}_q &=& \frac{id_{10}}{2m} \,\bar q \, [D^\mu, F_{\mu\nu}^+]
\, D^\nu \, q \, + \, {\, \rm h.c. \,} \, +  \cdots \,,\nonumber\\[2mm]
{\cal L}^{(4)}_q &=& \frac{e_6}{2} \, \la \chi_+ \ra \,
\bar q\,\sigma^{\mu\nu}\, F_{\mu\nu}^+ \, q\,
+\, \frac{e_7}{4} \,
\bar q\,\sigma^{\mu\nu}\, \{F_{\mu\nu}^+ \hat\chi_+\} \, q\,
+\, \frac{e_8}{2} \,
\bar q\,\sigma^{\mu\nu}\, \la F_{\mu\nu}^+ \hat\chi_+ \ra \, q\,
- \frac{e_{10}}{2}\,\bar q \, [D^\alpha, [D_\alpha, F_{\mu\nu}^+] ]
\sigma^{\mu\nu} \, q \, + \cdots, \nonumber
\en
where $\hat\chi_+ = \chi_+ - \frac{1}{3} \la{\chi_+}\ra\,$,
the symbols $\la \,\, \ra$, $[ \,\, ]$ and $\{ \,\, \}$
occurring in Eq.~(\ref{L_exp}) denotes the trace over flavor
matrices, commutator and anticommutator, respectively.
In Eq.~(\ref{L_exp}) we display only the terms involved
in the calculation of the dressed electromagnetic quark operator.
Here, for simplicity, we drop the contribution of vector mesons.
The detailed form of the chiral Lagrangian can be found
in Ref.~\cite{Faessler:2005gd}.

The couplings $m$ and $g$ denote the quark mass and axial charge
in the chiral limit, $c_i$, $d_i$ and $e_i$ are
the second-, third- and fourth-order low-energy
coupling constants, respectively, which encode the contributions of
heavy states. 
Parameter $m$ is counted as as quantity of order $O(1)$
in the chiral expansion. 

Here $q$ is the quark field, and the octet of pseudoscalar fields
\eq
\phi = \sum_{i=1}^{8} \phi_i\lambda_i = \sqrt{2}
\left(
\begin{array}{ccc}
\pi^0/\sqrt{2} + \eta/\sqrt{6}\,\, & \,\, \pi^+ \,\, & \, K^+ \\
\pi^- \,\, & \,\, -\pi^0/\sqrt{2}+\eta/\sqrt{6}\,\, & \, K^0\\
K^-\,\, & \,\, \bar K^0 \,\, & \, -2\eta/\sqrt{6}\\
\end{array}
\right)
\en
is contained in the SU(3) matrix
$U = u^2 = {\rm exp}(i\phi/F)$ where $F$ is the octet decay constant. 
We introduce the
standard notations~\cite{Gasser:1987rb,Becher:1999he,Fettes:1998ud}
\eq
& &D_\mu = \partial_\mu + \Gamma_\mu, \hspace*{.3cm}
\Gamma_\mu = \frac{1}{2} [u^\dagger, \partial_\mu u]
- \frac{i}{2} u^\dagger R_\mu u
- \frac{i}{2} u L_\mu u^\dagger, \\[2mm]
& &u_\mu = i u^\dagger \nabla_\mu U u^\dagger, \hspace*{.3cm}
\chi_\pm = u^\dagger \chi u^\dagger \pm u \chi^\dagger u, \hspace*{.3cm}
\chi = 2 B {\cal M} + \cdots \, .  \nonumber
\en
The fields $R_\mu$ and $L_\mu$ include external fields (electromagnetic
$A_\mu$, weak, etc.):
\eq
R_\mu \, = \, e \, Q \, A_\mu \, + \cdots \,, \hspace*{1cm}
L_\mu \, = \, e \, Q \, A_\mu \, + \cdots \nonumber
\en
where
$Q = {\rm diag} \{ 2/3,-1/3,-1/3 \}$ is the quark charge matrix.
The tensor $F_{\mu\nu}^+$ is defined as
$F_{\mu\nu}^+ \, = \, u^\dagger F_{\mu\nu}  Q u +
u F_{\mu\nu} Q u^\dagger$ where
$F_{\mu\nu} = \partial_\mu A_\nu - \partial_\nu A_\mu$ is the
conventional photon field strength tensor.
Here ${\cal M} = {\rm diag}\{\hat m, \hat m, \hat m_s\}$ is the mass
matrix of current quarks (we work in the isospin symmetry limit
with $\hat m_{u}= \hat m_{d}=\hat{m}=7$ MeV and the mass of
the strange quark $\hat m_s$ is related to the nonstrange one as
$\hat m_s = 25 \, \hat m$).

The quark vacuum condensate parameter is denoted by
\eq
B = - \frac{1}{F^2} \la 0|\bar u u|0 \ra  =
      - \frac{1}{F^2} \la 0|\bar d d|0 \ra  \,.
\en
To distinguish between
constituent and current quark masses we attach the symbol
$\ {\bf\hat{}}$ ("hat") when referring to the current quark masses.
We rely on the standard picture of chiral symmetry
breaking~($B \gg F$). In leading order of the chiral expansion
the masses of pseudoscalar mesons are given by
\eq\label{M_Masses}
M_{\pi}^2=2 \hat m B, \hspace*{.5cm}
M_{K}^2=(\hat m + \hat m_s) B, \hspace*{.5cm}
M_{\eta}^2= \frac{2}{3} (\hat m + 2 \hat m_s) B\,.
\en
In the numerical analysis we will use: $M_{\pi} = 139.57$ MeV,
$M_K = 493.677$ MeV (the charged pion and kaon masses),
$M_\eta = 574.75$ MeV and the canonical set of differentiated decay
constants:  $F_\pi = 92.4$ MeV, $F_K/F_\pi = 1.22$ and
$F_\eta/F_\pi = 1.3$~\cite{Gasser:1984gg}.

\subsection{Dressing of quark operators}

Any bare quark operator (both one- and two-body) can be dressed by
a cloud of pseudoscalar mesons and heavy states in a
straightforward manner by use of the effective chirally-invariant
Lagrangian ${\cal L}_{qU}$. To illustrate the idea of such a
dressing we consider the Fourier-transform of the electromagnetic
quark operator: \eq\label{bare_V} J_{\mu, \, {\rm em}}^{\rm
bare}(q) = \int d^4x \, e^{-iqx} \, j_{\mu, \, {\rm em}}^{\rm
bare}(x) \,, \quad \hspace*{.3cm} j_{\mu, {\rm em}}^{\rm bare}(x)
= \bar q(x) \, \gamma_\mu \, Q \, q(x)\,. \en In Fig.1 we display
the tree and loop diagrams which contribute to the dressed
electromagnetic operator $J_{\mu, \, {\rm em}}^{\rm dress}$ up to
fourth order. Note, here we restrict our consideration to the
one-body quark operator. An extension of our method to two-body
quark operators will be done in future.

The dressed quark operator $j_{\mu, \, {\rm em}}^{\rm dress}(x)$
and its Fourier transform $J_{\mu, \, {\rm em}}^{\rm dress}(q)$ have the
following forms
\eq\label{Jmu_dress}
j_{\mu, \, {\rm em}}^{\rm dress}(x) &=& \sum\limits_{q=u,d,s} \,
\biggl\{ f_D^q(-\partial^2) \, [ \bar q(x) \gamma_\mu q(x) ]
\, + \, \frac{f_P^q(-\partial^2)}{2m_q} \, \partial^\nu \,
[ \bar q(x) \sigma_{\mu\nu} q(x) ] \biggr\} \,\\
J_{\mu, \, {\rm em}}^{\rm dress}(q) &=& \int d^4x \, e^{-iqx} \,
j_{\mu, \, {\rm em}}^{\rm dress}(x) = \int d^4x \, e^{-iqx} \,
\sum\limits_{q=u,d,s}
\bar q(x) \, \biggl[ \, \gamma_\mu \, f_D^q(q^2) \, + \, \frac{i}{2m_q} \,
\sigma_{\mu\nu} \, q^\nu \, f_P^q(q^2) \, \biggr] \, q(x)\,, \nonumber
\en
where $m_q$ is the dressed constituent quark mass generated
by the chiral Lagrangian~(\ref{L_exp})
(see details in Ref.~\cite{Faessler:2005gd});
$f_D^u(q^2)$, $f_D^d(q^2)$, $f_D^s(q^2)$ and
$f_P^u(q^2)$, $f_P^d(q^2)$, $f_P^s(q^2)$ are
the Dirac and Pauli form factors of $u$, $d$ and $s$ quarks.
Here we use the appropriate sub- and superscripts with a definite
normalization of the set of $f_D^q(0) \, \equiv \, e_q$ (quark charges)
due to charge conservation. Note, that the dressed quark operator
satisfies current conservation:
$\partial^\mu \, j_{\mu, \, {\rm em}}^{\rm dress}(x) = 0$. 
Evaluation of the diagrams in Fig.1 is based on the {\it infrared
dimensional regularization} suggested in Ref.~\cite{Becher:1999he}
to guarantee a straightforward connection between loop and chiral
expansion in terms of quark masses and small external momenta.
We relegate the discussion of the calculational technique
to Ref.~\cite{Faessler:2005gd}.

To calculate the electromagnetic transitions between baryons we project
the dressed quark operator between the corresponding baryon states.
The master formula is:
\eq\label{master}
&&\la B(p^\prime) | \, J_{\mu, \, {\rm em}}^{\rm dress}(q)
\, | B(p) \ra \, = \, (2\pi)^4 \, \delta^4(p^\prime - p - q) \,
\bar u_B(p^\prime) \biggl\{ \gamma_\mu \, F_1^B(q^2) \, + \,
\frac{i}{2 \, m_B} \, \sigma_{\mu\nu} q^\nu
\, F_2^B(q^2) \biggr\} u_B(p) \, \nonumber\\
& = & (2\pi)^4 \, \delta^4(p^\prime - p - q) \sum\limits_{q = u,d,s}
\biggl\{f_D^q(q^2) \, \la B(p^\prime)|\,j_{\mu, q}^{\rm bare}(0)\,|B(p) \ra
+  i \, \frac{q^\nu}{2 \, m_q} \, f_P^q(q^2) \,
\la B(p^\prime)| \, j_{\mu\nu, q}^{\rm bare}(0) \, |B(p) \ra
\biggr\} \, ,
\en
where $B(p)$ and $u_B(p)$ are the baryon state and spinor,
respectively, normalized as
\eq
\la B(p^\prime) | B(p) \ra =
2 E_B \, (2\pi)^3 \, \delta^3(\vec{p}-\vec{p}^{\,\prime})\,, 
\hspace*{.5cm} 
\bar u_B(p) u_B(p) = 2 m_B
\en
with $E_B = \sqrt{m_B^2+\vec{p}^{\,2}}$ being the baryon energy
and $m_B$ the baryon mass.
In Eq.(\ref{master}) we focus on the diagonal $\frac{1}{2}^+
\to \frac{1}{2}^+$  transitions (the extension to the nondiagonal
transitions and transitions involving higher spin states like the
$\Delta(1232)$ isobar is straightforward).
Here $F_1^B(q^2)$ and $F_2^B(q^2)$ are the Dirac and Pauli baryon
form factors. In Eq.~(\ref{master}) we express the matrix elements of
the dressed quark operator in terms of the matrix elements of the
bare operators. In our application we deal with the bare quark
operators for vector $j_{\mu, q}^{\rm bare}(0)$ and tensor
$j_{\mu\nu, q}^{\rm bare}(0)$ currents defined as
\eq\label{bare_operators}
j_{\mu, q}^{\rm bare}(0) \, = \, \bar q(0) \, \gamma_\mu \, q(0)\,,
\hspace*{1cm} j_{\mu\nu, q}^{\rm bare}(0) \, = \, \bar q(0) \,
\sigma_{\mu\nu} \, q(0)\,.
\en
\noindent
Eq.~(\ref{master}) contains our main result: we perform a model-independent
factorization of the effects of hadronization and confinement contained in
the matrix elements of the bare quark operators $j_{\mu, q}^{\rm bare}(0)$
and $j_{\mu\nu, q}^{\rm bare}(0)$ and the effects dictated by chiral
symmetry (or chiral dynamics) which are encoded in the relativistic form
factors $f_D^q(q^2)$ and $f_P^q(q^2)$. Due to this factorization
the calculation of $f_D^q(q^2)$ and $f_P^q(q^2)$, on one side,
and the matrix elements of $j_{\mu, q}^{\rm bare}(0)$ and
$j_{\mu\nu, q}^{\rm bare}(0)$, on the other side, can be done
independently. In particular, in a first step we derived
a model-independent formalism based on the ChPT Lagrangian,
which is formulated in terms of constituent quark degrees of freedom,
for the calculation of $f_D^q(q^2)$
and $f_P^q(q^2)$. The calculation of the matrix elements of the bare
quark operators can then be relegated to quark models based on specific
assumptions about hadronization and confinement. The explicit forms of
$f_D^q(q^2)$ and $f_P^q(q^2)$ are given in Appendix~C of
Ref.~\cite{Faessler:2005gd}.

\subsection{Matching to ChPT}

The matrix elements of the bare quark operators should be
calculated using specific model-dependent assumptions about
hadronization and confinement. In Ref.~\cite{Faessler:2005gd} it
was shown that in the case of nucleons the use of certain symmetry
constraints leads to a set of relationships between the nucleon
and corresponding quark form factors at zero momentum transfer. In
general, due to Lorentz and gauge invariance, the matrix elements
in Eq.~(\ref{master}) can be written as \eq\label{val_expansion}
\la B(p^\prime) | \, j_{\mu, q}^{\rm bare}(0) \, | B(p) \ra &=&
\bar u_B(p^\prime) \biggl\{ \gamma_\mu \, F_1^{Bq}(q^2) \, + \,
\frac{i}{2 \, m_B}
\, \sigma_{\mu\nu} \, q^\nu \,  F_2^{Bq}(q^2) \biggr\} u_B(p)\,, \\
i \, \frac{q^\nu}{2 \, m_q}
\la B(p^\prime) | \, j_{\mu\nu, q}^{\rm bare}(0) \, | B(p) \ra &=&
\bar u_B(p^\prime) \biggl\{ \gamma_\mu \, G_1^{Bq}(q^2)
\, + \, \frac{i}{2 \, m_B}
\, \sigma_{\mu\nu} \, q^\nu \,  G_2^{Bq}(q^2) \biggr\} u_B(p) \, ,
\nonumber
\en
where $F_{1(2)}^{Bq}(q^2)$ and $G_{1(2)}^{Bq}(q^2)$ are the Pauli and
Dirac form factors describing the distribution of quarks of flavor
$q=u, d, s$ in the baryon $B$.

Let us briefly review the constraints on the nucleon form factors
derived in Ref.~\cite{Faessler:2005gd}. The first set of relations
arise from charge conservation and isospin invariance:
\eq
& &F_1^{pu}(0) = F_1^{nd}(0) = 2\,, \hspace*{.5cm}
F_1^{pd}(0) = F_1^{nu}(0) = 1\,, \hspace*{.5cm}
G_1^{Nq}(0) = 0\,, \\
& &F_2^{pu}(0) = F_2^{nd}(0)\,, \hspace*{.5cm}
F_2^{pd}(0) = F_2^{nu}(0)\,, \hspace*{.5cm}
G_2^{pu}(0) = G_2^{nd}(0)\,, \hspace*{.5cm}
G_2^{pd}(0) = G_2^{nu}(0)\,. \nonumber
\en
Note, that the quantities $G_2^{Nq}(0)$ are related to the bare
nucleon tensor charges $\delta_{Nq}^{\rm bare}$~\cite{Faessler:2005gd}. 
The second set of constraints are the
so-called {\it chiral symmetry constraints}. They are dictated by
the infrared-singular structure of QCD and reproduce the leading
nonanalytic (LNA) contributions to the magnetic moments and the
charge and magnetic radii of
nucleons~\cite{Kubis:2000zd,Beg:1973sc}: 
\eq\label{chiral_constr}
\mu_p &=& - \frac{g_A^2}{8 \pi} \, \frac{M_\pi}{F_\pi^2} \,
\mnz \, + \, \cdots \,, \nonumber\\
\la r^2 \ra^E_p &=& - \frac{1 + 5 g_A^2}{16 \, \pi^2 \, F_\pi^2} \,
{\rm ln}\frac{M_\pi}{\mnz} \, + \, \cdots \,, \\
\la r^2 \ra^M_p &=& \frac{g_A^2}{16 \, \pi \, F_\pi^2 \, \mu_p} \,
\frac{\mnz}{M_\pi} \, + \, \cdots \,,  \nonumber 
\en 
where $g_A$
and $\mnz$ are the axial charge and the mass of the nucleon in the
chiral limit. In order to fulfill the strictures of
chiral symmetry~(\ref{chiral_constr}) we demand the following
identities involving the $F_2^{Nq}(0)$ and $G_2^{Nq}(0)$ form
factors 
\eq\label{chiral_constr2} & &1 + F_2^{pu}(0) - F_2^{pd}(0)
= G_2^{pu}(0) - G_2^{pd}(0) =
\biggl(\frac{g_A}{g}\biggr)^2\, \frac{m_N}{\bar m}\,, \\
& &1 + F_2^{nd}(0) - F_2^{nu}(0) = G_2^{nd}(0) - G_2^{nu}(0) =
\biggl(\frac{g_A}{g}\biggr)^2 \, \frac{m_N}{\bar m} \, 
\en 
where $\bar m = m_u = m_d$ is the dressed nonstrange constituent 
quark mass in the isospin limit. 
In Ref.~\cite{Faessler:2005gd}, the SU(6)-symmetry
relations of the naive nonrelativistic quark model have been 
used for further constraints on $F_2^{Ni}(0)$
and $G_2^{Ni}(0)$. 
In this paper we go beyond the simple SU(6) picture, utilizing
the relativistic constituent quark
model~\cite{Ivanov:1996pz}-\cite{Faessler:2006ft}
to calculate the bare baryonic matrix elements or to evaluate
the contribution from the valence degrees of freedom.

\subsection{Evaluation of the matrix elements of
the valence quark operators}

In this section we discuss the calculation of the baryonic matrix
elements \eq\label{matrix_VT} \la B(p^\prime)|\,j_{\mu, q}^{\rm
bare}(0)\,|B(p) \ra\, \hspace*{.5cm} {\rm and} \hspace*{.5cm} \la
B(p^\prime)| \, j_{\mu\nu, q}^{\rm bare}(0) \, |B(p) \ra \en
induced, respectively, by the bare quark operators:
\eq\label{bare_operators2} j_{\mu, q}^{\rm bare}(0) \, = \, \bar
q(0) \, \gamma_\mu \, q(0)\,, \hspace*{.5cm} {\rm and}
\hspace*{.5cm} j_{\mu\nu, q}^{\rm bare}(0) \, = \, \bar q(0) \,
\sigma_{\mu\nu} \, q(0)\,. \en \noindent We will consistently
employ the relativistic three-quark model
(RQM)~\cite{Ivanov:1996pz}-\cite{Faessler:2006ft} to compute such
matrix elements~(\ref{matrix_VT}). The RQM was previously
successfully applied for the study of properties of baryons
containing light and heavy
quarks~\cite{Ivanov:1996fj}-\cite{Faessler:2006ft}. The main
advantages of this approach are: Lorentz and gauge invariance, a
small number of parameters, and modelling of effects of strong
interactions at large ($\sim 1$ fm) distances. Various properties
of light and heavy baryons have been analyzed within this
RQM~\cite{Ivanov:1996pz}-\cite{Faessler:2006ft}, and a preliminary
analysis of the electromagnetic properties of nucleons has been
performed in Ref.~\cite{Ivanov:1996pz} where the effects of
valence quarks have been consistently taken into account. 
Here we extend this analysis to the
case of hyperons as well as to the $N \to \Delta\gamma$
transitions and we include meson-cloud effects.

Let us begin by briefly reviewing the basic notions of the RQM
approach~\cite{Ivanov:1996pz}-\cite{Faessler:2006ft}. The RQM is
essentially based on an interaction Lagrangian describing the
coupling between baryons and their constituent quarks. The
coupling of a baryon $B(q_1q_2q_3)$ to its constituent quarks
$q_1$, $q_2 $ and $q_3$ is described by the Lagrangian
\eq\label{Lagr_str} {\cal L}_{\rm int}^{\rm str}(x) = g_B \bar
B(x) \, \int\!\! dx_1 \!\! \int\!\! dx_2 \!\! \int\!\! dx_3 \,
F_B(x,x_1,x_2,x_3) \, J_B(x_1,x_2,x_3) \, + \, {\rm H.c.} \en
where $J_{B}(x_1,x_2,x_3)$ is the three-quark current with the
quantum numbers of the relevant baryon
$B$~\cite{Ioffe:1982ce,Efimov:1987na}. One has 
\eq
J_{B}(x_1,x_2,x_3) \, = \, \epsilon^{a_1a_2a_3} \, \Gamma_1 \,
q^{a_1}_1(x_1) \, q^{a_2}_2(x_2) C \, \Gamma_2 \, q^{a_3}_3(x_3)
\, , 
\en 
where $\Gamma_{1,2}$ are Dirac structures,
$C=\gamma^{0}\gamma^{2}$ is the charge conjugation matrix and
$a_i, i=1,2,3$ are color indices.

The function $F_B$ is related to the scalar part of the
Bethe-Salpeter amplitude and characterizes the finite size
of the baryon.
In the following we use a particular form for the
vertex function~\cite{Ivanov:1996pz}-\cite{Faessler:2006ft}
\eq\label{vertex}
F_B(x,x_1,x_2,x_3) \, = \, \delta^4(x - \sum\limits_{i=1}^3 w_i x_i) \;
\Phi_B\biggl(\sum_{i<j}( x_i - x_j )^2 \biggr)
\en
where $\Phi_B$ is the correlation function of three constituent
quarks with masses $m_1$, $m_2$, $m_3$. The variable $w_i$ is defined by
$w_i=m_i/(m_1+m_2+m_3)$
and therefore depends only on the relative Jacobi coordinates $(\xi_{1},
\xi_{2})$ as $\Phi_{B}(\xi^2_1+\xi^2_2)$, where
\eq\label{coordinates}
x_1&=&x \, - \, \frac{\xi_1}{\sqrt{2}} \, (w_2+w_3)
        \, + \, \frac{\xi_2}{\sqrt{6}} \, (w_2 - w_3)\,,\nonumber\\
x_2&=&x \, + \, \frac{\xi_1}{\sqrt{2}} \,  w_1
        \, - \, \frac{\xi_2}{\sqrt{6}} \, (w_1 + 2w_3)\,, \\
x_3&=&x \, + \, \frac{\xi_1}{\sqrt{2}} \,  w_1
        \, + \, \frac{\xi_2}{\sqrt{6}} \, (w_1 + 2w_2)\,,\nonumber
\en
and with $x \,= \, \sum\limits_{i=1}^3 w_i x_i$ being
the center of mass (CM) coordinate.
Expressed in relative Jacobi coordinates and the center of mass
coordinate, the Fourier transform of the vertex function
reads~\cite{Ivanov:1996pz}-\cite{Faessler:2006ft}:
\eq
\Phi_{B}(\xi^2_1+\xi^2_2)=
\int\!\frac{d^4p_1}{(2\pi)^4}\!\int\!\frac{d^4p_2}{(2\pi)^4}
e^{-ip_1\xi_1-ip_2\xi_2}{\widetilde{\Phi}}_{B}(-p^2_1-p^2_2) \,.
\en
The choice of light baryon three-quark currents has been
discussed in detail in Refs.~\cite{Ioffe:1982ce,Efimov:1987na}.
(For the discussion of the heavy baryon currents
see Refs.~\cite{Shuryak:1980pg}
and~\cite{Ivanov:1996pz}-\cite{Faessler:2006ft}).
When restricted to the unitary flavor SU(3) symmetry
and the octet of light baryons, one can construct
two linearly independent currents: vector and tensor.
For the light baryon decuplet there exists only a single
vector current.  In Appendix~A we list these three-quark currents
for the baryon octet and for the $\Delta(1232)$-isobar.
Note that the vector and tensor currents of the baryon
octet~\cite{Ioffe:1982ce,Efimov:1987na} are degenerate
in the nonrelativistic
limit. In the following we show that even in the relativistic case
they give similar predictions for the magnetic moments of the
nucleons, hyperons and for the $N\to\Delta\gamma$
transition.  The quantities which {\it are} sensitive to the
choice of the baryon octet currents are the $E2/M1$ and $C2/M1$
ratios, which are generated by relativistic effects and vanish in
the nonrelativistic limit.

The baryon-quark coupling constants $g_B$ are determined by the
compositeness
condition~\cite{Ivanov:1996pz}-\cite{Faessler:2006ft} (see
also~\cite{Weinberg:1962hj,Efimov:1993ei}), which implies that the
renormalization constant of the hadron wave function is set equal
to zero: \eq Z_B = 1 - \Sigma^\prime_B(m_B) = 0 \, \en where
$\Sigma^\prime_B (m_B) = g_B^2 \Pi^\prime_B(m_B)$ is the
derivative of the baryon mass operator described by the diagram in
Fig.2 and $m_B$ is the baryon mass. To clarify the physical
meaning of this condition, we first want to remind the reader that
the renormalization constant $Z_B^{1/2}$ can also be interpreted
as the matrix element between the physical and the corresponding
bare state. For $Z_B=0$ it then follows that the physical state
does not contain the bare one and is described as a bound state.
The interaction Lagrangian Eq.~(\ref{Lagr_str}) and the
corresponding free parts describe both the constituents (quarks)
and the physical particles (hadrons) which are taken to be the
bound states of the constituents. As a result of the interaction,
the physical particle is dressed, {\it i.e.} its mass and its wave
function have to be renormalized. The condition $Z_B=0$ also
effectively excludes the constituent degrees of freedom from the
physical space and thereby guarantees that there is no double
counting for the physical observable under consideration. In this
picture the constituents exist in virtual states only. One of the
corollaries of the compositeness condition is the absence of a
direct interaction of the dressed charged particle with the
electromagnetic field. Taking into account both the tree-level
diagram and the diagrams with the self-energy and counter-terms
insertions into the external legs (that is the tree-level diagram
times $(Z_B -1$)) one obtains a common factor $Z_B$  which is
equal to zero~\cite{Efimov:1993ei}.

The quantities of interest, the matrix elements~(\ref{matrix_VT}),
are described by the triangle diagram in Fig.3a. In the case of
the matrix element of the vector current we need to take into account
two additional so-called ``bubble'' diagrams in Figs.3b and 3c
to guarantee gauge invariance of matrix elements (see details
in Refs.~\cite{Ivanov:1996pz}-\cite{Faessler:2006ft}
and~\cite{Mandelstam:1962mi}). In particular, the ``bubble'' diagrams
are generated by the non-local coupling of the baryon to the
constituent quarks and the external gauge field which arises after
gauging of the non-local strong interaction Lagrangian~(\ref{Lagr_str})
containing the vertex function~(\ref{vertex}).

In the evaluation of the quark-loop diagrams we use
the free fermion propagator for the constituent
quark~\cite{Ivanov:1996pz}-\cite{Faessler:2006ft}:
\eq\label{quark_propagator}
i \, S_q(x-y) = \langle 0 | T \, q(x) \, \bar q(y)  | 0 \rangle
\ = \ \int\frac{d^4k}{(2\pi)^4i} \, e^{-ik(x-y)} \ \tilde S_q(k)
\en
where
\eq
\tilde S_q(k) = \frac{1}{m_q-\not\! k -i\epsilon}
\en
is the usual free fermion propagator in momentum space.
We shall avoid the appearance of unphysical imaginary parts
in Feynman diagrams by postulating the condition that
the baryon mass must be less than the sum of the constituent
quark masses $M_{B} < \sum_{i}m_{q_{i}}$.

In the next step we have to specify the vertex function $\tilde\Phi_B$,
which characterizes the finite size of the baryons.
In principle, its functional form can be calculated from the solutions
of the Bethe-Salpeter equation for baryon bound
states~\cite{Ivanov:1998ya,Alkofer:2004yf}.
In Refs.~\cite{Anikin:1995cf} it was found that, using various
forms for the vertex function, the basic hadron observables
are insensitive to the details of
the functional form of the hadron-quark vertex form factor.
We will use this observation as a guiding principle and choose a simple
Gaussian form for the vertex function $\tilde\Phi_B$.
Any choice for $\tilde\Phi_B$ is appropriate
as long as it falls off sufficiently fast in the ultraviolet region of
Euclidean space to render the Feynman diagrams ultraviolet finite.
We employ the Gaussian form
\eq\label{Gauss_CF}
\tilde\Phi_B(k_{1E}^2,k_{2E}^2  )
\doteq \exp( - [k_{1E}^2 + k_{2E}^2]/\Lambda^2_B )\,,
\en
for the vertex function, where $k_{1E}$ and $k_{2E}$ are the
Euclidean momenta. Here $\Lambda_{B}$ is a size parameter, which
parametrizes the distribution of quarks inside a given baryon. In
previous papers we determined the following set of parameters for
light baryons:
\eq
m_u = m_d = 420  \ \ {\rm MeV}\,,  \ \
m_s = 570  \ \ {\rm MeV}\,,  \ \
\Lambda_B = 1.25 \ \ {\rm GeV}\,.
\en
Note that the quoted value of the nonstrange constituent quark mass
and the size parameter $\Lambda_B$ have been obtained from the
analysis of nucleon properties using the tensor $3q$ current,
with the inclusion of only valence degrees of freedom.
Below we intend to test the second choice -- the vector
current -- and also include the meson-cloud contributions to
the baryon properties.

\subsection{$N \to \Delta \gamma$ transition}

In this subsection we specify our approach for the case
of the $N \to \Delta \gamma$ transition. In particular, we discuss
in detail two important issues: i)~projection of the dressed quark
operator between nucleon and $\Delta(1232)$ states and ii)~evaluation
of the bare vector and tensor valence quark operators between the
nucleon and the $\Delta(1232)$.

The projection of the dressed quark operator between nucleon and
$\Delta(1232)$ states reads
\eq\label{master3}
&&\la \Delta(p^\prime) | \, J_{\mu, \, {\rm em}}^{\rm dress}(q)
\, | N(p) \ra \, = \, (2\pi)^4 \, \delta^4(p^\prime - p - q) \,
\bar u_\Delta^\nu(p^\prime) \, \Lambda_{\mu\nu}(p,p^\prime) \, u_N(p) \,
\nonumber\\
& = & (2\pi)^4 \, \delta^4(p^\prime - p - q) \sum\limits_{q = u,d}
\biggl\{f_D^q(q^2) \, \la \Delta(p^\prime)|\, j_{\mu, q}^{\rm
bare}(0)\,|N(p) \ra +  i \, \frac{q^\nu}{2 \, m_q} \, f_P^q(q^2)
\, \la \Delta(p^\prime)| \, j_{\mu\nu, q}^{\rm bare}(0) \, |N(p)
\ra \biggr\} \, , \en where $\Lambda_{\mu\nu}(p,p^\prime)$ is the
$N \to \Delta  \gamma$ vertex function, and $\bar
u_\Delta^\nu(p^\prime)$ is the spin-$\frac{3}{2}$ Rarita-Schwinger
spinor satisfying the supplementary
conditions~\cite{Rarita:1941mf}: \eq\label{Rarita_Schwinger} \bar
u_\Delta^\nu(p^\prime) \, \gamma_\nu = 0 \hspace*{.5cm} {\rm and}
\hspace*{.5cm} \bar u_\Delta^\nu(p^\prime) \, p^{\prime}_\nu =
0 \,. \en The vertex function $\Lambda_{\mu\nu}(p,p^\prime)$ for
on-shell nucleon and $\Delta$-isobar states can be decomposed in
terms of relativistic form factors $b_i(q^2)$ with $i=1,2,3,4$:
\eq\label{set_b} \Lambda_{\mu\nu}(p,p^{\prime}) = [ g_{\mu\nu}
b_1(q^2) + p_\mu q_\nu  b_2(q^2) + \gamma_\mu q_\nu  b_3(q^2) +
q_\mu q_\nu  b_4(q^2) ] \gamma^5 \en Due to gauge invariance the
fourth form-factor is a linear combination of the other three: 
\eq
b_1(q^2) + b_2(q^2) p\cdot q + b_3(q^2) m_+ = - q^2 b_4(q^2) \,,
\en 
where $m_\Delta = 1232$ MeV is the mass of the
$\Delta$-isobar, $p\cdot q = (m_+ m_- - q^2)/2$ and $m_\pm =
m_\Delta \pm m_N$. The $N\Delta$ vertex function
$\Lambda_{\mu\nu}(p,p^{\prime})$ can then be rewritten in a
manifestly gauge-invariant form in terms of $b_1$, $b_2$ and
$b_3$: 
\eq \Lambda_{\mu\nu}(p,p^{\prime}) \, =  \, [
g_{\mu\nu}^\perp b_1(q^2) + p_\mu^\perp q_\nu  b_2(q^2) +
\gamma_\mu^\perp q_\nu b_3(q^2) ] \gamma^5 \en or in terms of
$b_2$, $b_3$ and $b_4$: \eq\label{set_b2b3b4}
\Lambda_{\mu\nu}(p,p^{\prime}) \, =  \, \biggl[ L_{2 \,
\mu\nu}^\perp \, b_2(q^2) \, + \, L_{3 \, \mu\nu}^\perp  \,
b_3(q^2) \, + \, L_{4 \, \mu\nu}^\perp \, b_4(q^2) \biggr]
\gamma^5 
\en 
where the superscript $\perp$ denotes the
Lorentz-structures perpendicular to the photon momentum: \eq &
&g_{\mu\nu}^\perp = g_{\mu\nu} - \frac{q_\mu q_\nu}{q^2}\,,
\hspace*{.5cm} p^\perp_\mu = p_\mu -  q_\mu \frac{pq}{q^2}\,,
\hspace*{.5cm}
\gamma^\perp_\mu = \gamma_\mu - q_\mu \frac{\not\! q}{q^2} \,, \\
& &L_{2 \, \mu\nu}^\perp \, = \,
p_\mu^\perp q_\nu  - g_{\mu\nu}^\perp \, pq \,,
\hspace*{.5cm}
L_{3 \, \mu\nu}^\perp \, = \,
\gamma_\mu^\perp q_\nu  - g_{\mu\nu}^\perp \, m_- \,, \hspace*{.5cm}
L_{4 \, \mu\nu}^\perp \, = \,  - g_{\mu\nu}^\perp \,, \nonumber
\en
with
\eq
g_{\mu\nu}^\perp \, q^\mu = 0\,, \hspace*{.5cm}
p^\perp_\mu \, q^\mu = 0\,, \hspace*{.5cm}
\gamma^\perp_\mu \, q^\mu = 0\,.
\en
It is easy to see that the gauge invariance of the $N \to \Delta\gamma$
matrix element is fulfilled:
$q^\mu \, \Lambda_{\mu\nu}(p,p^\prime) = 0$.
Alternative but equivalent sets of relativistic form
factors defining the $N\to\Delta\gamma$
transition~\cite{Bjorken:1966ij}-\cite{Braun:2005be} are given
in Appendix~B.

Note that for the evaluation of the $N \to \Delta \gamma$ matrix
element we use the {\it same} universal dressed electromagnetic
quark operators including chiral corrections~(\ref{Jmu_dress}).
For the calculation of the bare matrix elements $\la
\Delta(p^\prime)|\, j_{\mu, q}^{\rm bare}(0)\,|N(p) \ra$ and $\la
\Delta(p^\prime)| \, j_{\mu\nu, q}^{\rm bare}(0) \, |N(p) \ra$ we
apply the same quark approach
RQM~\cite{Ivanov:1996pz}-\cite{Faessler:2006ft} as for the case of
the octet transitions. Again, for the case of the ``vector''
matrix element $\la \Delta(p^\prime)|\, j_{\mu, q}^{\rm
bare}(0)\,|N(p) \ra$ we need to take into account the triangle
diagram in Fig.3a as well as the two ``bubble'' diagrams shown in
Figs.3b and 3c. For the case of the ``tensor'' matrix element $\la
\Delta(p^\prime)| \, j_{\mu\nu, q}^{\rm bare}(0) \, |N(p) \ra$ we
require the contribution of the triangle diagram only. In
addition, due to the nondiagonality of the $N \to \Delta \gamma$
transition, we need to include the diagram in Fig.3d in the
calculation of the ``vector'' matrix element in order to guarantee
gauge invariance.  This diagram describes the sub-process wherein
the nucleon converts into the $\Delta$-isobar via a quark loop
followed by the interaction of the $\Delta$ with the external
field.  Note that the analogous diagram where the nucleon
interacts with the external field and then converts into the
$\Delta$ vanishes due to the Rarita-Schwinger
conditions~(\ref{Rarita_Schwinger}).

In analogy with the $\frac{1}{2}^+ \to \frac{1}{2}^+$ transitions
[see Eq.~(\ref{val_expansion})] we, for convenience, perform the
expansion
of the bare matrix elements describing $N \to \Delta$ transitions:
\eq
\la \Delta(p^\prime)|\, j_{\mu, q}^{\rm bare}(0)\,|N(p) \ra
&=& \bar u_\Delta^\nu(p^\prime) \, \sum\limits_{i=2}^4 \,
L_{i \, \mu\nu}^\perp \, b_i^V(q^2) \, \gamma^5 \, u_N(p)\,, \\
i \, \frac{q^\nu}{2 \, m_q}  \,
\la \Delta(p^\prime)| \, j_{\mu\nu, q}^{\rm bare}(0) \, |N(p) \ra
&=& \bar u_\Delta^\nu(p^\prime) \ \, \sum\limits_{i=2}^4 \,
L_{i \, \mu\nu}^\perp \, b_i^T(q^2) \, \gamma^5 \, u_N(p)\,,
\en
where the superscripts $V$ and $T$ denote the partial contributions
of vector and tensor matrix element to the relativistic form
factors $b_i$.  Finally, the total results for the form factors
$b_i$ are:
\eq
& &b_i(q^2) \, = \, b_i^{\rm bare}(q^2) \, + \, b_i^{\rm cloud}(q^2) \,,
\nonumber\\ [2mm]
& &b_i^{\rm bare}(q^2) \, = \,
\sum\limits_{q = u,d} \, e_q   \, b_i^V(q^2)\,,
\hspace*{.5cm}
   b_i^{\rm cloud}(q^2) \, = \,
\sum\limits_{q = u,d} \biggl[
 ( f_D^q(q^2) - e_q )  \,  b_i^V(q^2) \,
+ \, f_P^q(q^2)  \, b_i^T(q^2) \biggr] \,. \en where we have
separated each form factor into its bare and meson cloud
components.

\newpage 

\section{Physical applications}

In this section we consider the application of our technique to
the problem of magnetic moments of light baryons and the static
characteristics of the $N \to \Delta \gamma$ transition.  We
calculate the contributions of both valence and sea-quarks to
these quantities using the approach discussed above.  We remind
the reader that such an analysis was performed in
Ref.~\cite{Faessler:2005gd} using symmetry constraints in order to
determine values of valence baryon form factors at zero recoil. In
particular, exact values of the contributions of the valence
degrees of freedom to the Pauli form factors were deduced using
the requirements of SU(6) spin-flavor symmetry.  In addition we
considered a second possibility when we included SU(6) breaking
corrections but without specific calculations.  Here we precisely
evaluate the valence quark effects (matrix elements of the bare
quark operators~(\ref{matrix_VT})) using a Lorentz covariant
framework, which helps to take relativistic effects into account.
The proper inclusion is essential for a consistent calculation of
the $N \to \Delta \gamma$ transition. In this paper we restrict
our attention to the magnetic moments of the baryon octet and the
multipoles of the $N \to \Delta \gamma$ transition.

\subsection{Definition of baryon quantities}

Below we give a set of definitions of baryon quantities which are
the subject of the present calculations. First we recall the
definition of the magnetic moments $\mu_B$ of the baryon octet in
terms of the Dirac -- $F_1^B$ -- and Pauli - $F_2^B$ -- form
factors derived in Eq.~(\ref{master}): \eq \mu_B \, = \, [ \,
F_1^B(0) + F_2^B(0) \, ] \,\, \frac{e}{2 m_B} \,, \en where we
have set $\hbar = 1$. In terms of the nuclear magneton -- $\mu_N =
\frac{e \, \hbar}{2 m_p}$ -- the baryon magnetic moment is given
by \eq \mu_B  = \ [ \, F_1^B(0) + F_2^B(0) \, ] \,\,
\frac{m_p}{m_B} \,, \en where $m_p$ is the proton mass.  The
off-diagonal $M1$ moment $\mu_{\Sigma\Lambda}$ defining the
transition $\Sigma^0\to\Lambda \gamma$ is given (again in units of
nuclear magnetons) \eq \mu_{\Sigma\Lambda} =
F_2^{\Sigma\Lambda}(0) \, \frac{2m_p}{m_\Sigma + m_\Lambda} \,,
\en where $F_2^{\Sigma\Lambda}(0)$ is the value of the
corresponding Pauli form factor at zero recoil.

As noted above, in our formalism the magnetic moments of the octet
baryon can be split into the contribution from valence quarks
$\mu_B^{\rm bare}$ and from the meson cloud $\mu_B^{\rm cloud}$:
\eq\label{mu_total}  
\mu_B \, = \, \mu_B^{\rm bare} \, + \, \mu_B^{\rm cloud} 
\en
where 
\eq 
\mu_B^{\rm bare} &=& \sum_{q=u,d,s} f_D^q(0) \, \left(
F_1^{Bq}(0)+F_2^{Bq}(0) \right) \,,
\label{mu_bare}\\
\mu_B^{\rm cloud} &=& \sum_{q=u,d,s} f_P^q(0)
G_2^{Bq}(0) \,. \label{mu_cloud}
\en
Here the values of the meson-cloud Dirac form factors $f_D^q$ at
zero recoil coincide with the quark charges due to charge
conservation: $f_D^q(0) \equiv e_q$.  These meson-cloud
form factors $f_D^q$ and $f_P^q$ have been calculated in
Ref.~\cite{Faessler:2005gd}, and the calculational method for the
valence-quark form factors $F_i^{Bq}$ and $G_i^{Bq}$ has been
discussed in detail in Refs.~\cite{Ivanov:1996pz}-\cite{Faessler:2006ft}.

A complete description of the $N \to \Delta \gamma$ transition is
then given in terms of the set of relativistic form factors
$b_i(q^2)$ (expressions in terms of form factors from equivalent
definitions can be found using the relations given in Appendix~B
and in Refs.~\cite{Bjorken:1966ij}-\cite{Braun:2005be}):

1) Magnetic form factor $G_{M1}(Q^2)$:

\eq
G_{M1}(Q^2) \, = \, \frac{1}{4}
\biggl\{
b_3(Q^2) \frac{m_+ (3m_\Delta + m_N) + Q^2}{m_\Delta} \, + \,
b_2(Q^2) ( m_+ m_- + Q^2 ) \, - \, 2 b_4(Q^2) Q^2
\biggr\} \,.
\en

2) Electric form factor $G_{E2}(Q^2)$:

\eq
G_{E2}(Q^2) \, = \, \frac{1}{4}
\biggl\{
b_3(Q^2) \frac{m_+ m_- - Q^2}{m_\Delta} \, + \,
b_2(Q^2) ( m_+ m_- + Q^2 ) \, - \, 2 b_4(Q^2) Q^2
\biggr\} \,.
\en

3) Coulombic form factor $G_{C2}(Q^2)$:

\eq
G_{C2}(Q^2) \, = \, \frac{|\vec q \,|}{2}
\biggl\{
b_3(Q^2) \, + \, b_2(Q^2) E_N  \, + \, b_4(Q^2) \omega
\biggr\} \,.
\en

4) Helicity amplitudes $A_{3/2}(Q^2)$ and $A_{1/2}(Q^2)$:

\eq
A_{3/2}(Q^2) &=& - \, \sqrt{\frac{\pi \, \alpha \, \omega}{2 m_N^2}} \,\,
[ G_{M1}(Q^2) \, + \, G_{E2}(Q^2) ] \,, \\
A_{1/2}(Q^2) &=& - \, \sqrt{\frac{\pi \, \alpha \, \omega}{6 m_N^2}} \,\,
[ G_{M1}(Q^2) \, - \, 3 \, G_{E2}(Q^2) ] \,.
\en

5) Ratios EMR = $E2/M1 = - G_{E2}(Q^2)/G_{M1}(Q^2)$ and
          CMR = $C2/M1 = - G_{C2}(Q^2)/G_{M1}(Q^2)$:

\eq
{\rm EMR}(Q^2) \, = \, - \, \frac{G_{E2}(Q^2)}{G_{M1}(Q^2)}
\hspace*{.5cm} {\rm and} \hspace*{.5cm}
{\rm CMR}(Q^2) \, = \,  - \,  \frac{G_{C2}(Q^2)}{G_{M1}(Q^2)} \,.
\en

6) Transition dipole moment $\mu_{N\Delta}$:

\eq
\mu_{N\Delta} \, = \, \frac{2}{\sqrt{6}} \, G_{M1}(0) \,.
\en

7) Transition quadrupole moment $Q_{N\Delta}$:

\eq
Q_{N\Delta} \, = - \, \frac{4 \, \sqrt{6}}{m_+ \, m_-} \,
\frac{m_\Delta}{m_N} \, G_{E2}(0) \,.
\en

8) $\Delta^+ \to p + \gamma$ decay width:

\eq
\Gamma(\Delta^+ \to p \gamma) = \frac{m_\Delta \, m_N}{8 \, \pi} \,
\biggl[ 1 - \frac{m_N^2}{m_\Delta^2}\biggr]^2
 \, \biggl\{ \, |A_{1/2}(0)|^2 + |A_{3/2}(0)|^2 \biggr\}\,,
\en
where $Q^2 = - q^2$ is an Euclidean momentum squared,
$\alpha = 1/137$ is the fine structure coupling,
\eq
E_N \, = \, m_\Delta - \omega = \frac{m_\Delta^2 + m_N^2 + Q^2}{2 m_\Delta}
\hspace*{.5cm} {\rm and} \hspace*{.5cm}
\omega = \frac{m_\Delta^2 - m_N^2 - Q^2}{2 m_\Delta}
\en
are the nucleon and photon energies, and
\eq
|\vec q \,| \, = \,
\frac{\lambda^{1/2}(m_\Delta^2,m_N^2,-Q^2)}{2 m_\Delta}
\en
is the 3-momentum of the virtual photon in the
$\Delta$-isobar rest frame.
Here
\eq
\lambda(x,y,z) \, = \, x^2 + y^2 + z^2 - 2 xy - 2 xz - 2 yz
\en
is the K\"{a}llen triangle function.

Note, that the form factors $G_{M1}(Q^2)$ and $G_{E2}(Q^2)$ can
be written in a more compact form
as combinations of two form factors $b_1(Q^2)$ and $b_3(Q^2)$
using the identity (39):
\eq
G_{M1}(Q^2) \, = \, \frac{1}{2}
\biggl\{ - b_1(Q^2) + b_3(Q^2)
\frac{m_+^2 + Q^2}{2m_\Delta} \biggr\} \,.
\en
and
\eq
G_{E2}(Q^2) \, = \, \frac{1}{2}
\biggl\{ - b_1(Q^2) - b_3(Q^2)
\frac{m_+^2 + Q^2}{2m_\Delta} \biggr\} \,.
\en
Therefore, the sum of $G_{M1}(Q^2)$ and $G_{E2}(Q^2)$ is defined
by the $b_1(Q^2)$ form factor, while their difference involves
only the form factor $b_3(Q^2)$:
\eq
G_{M1}(Q^2) + G_{E2}(Q^2) &=& - b_1(Q^2) \,, \nonumber\\
G_{M1}(Q^2) - G_{E2}(Q^2) &=&   b_3(Q^2) \,
\frac{m_+^2 + Q^2}{2m_\Delta} \,.
\en

\subsection{Numerical results}

As stressed above, for the octet states there exist two possible
choices for the three-quark current: vector and tensor. A
preliminary analysis (see also Ref.~\cite{Ivanov:1996pz}) showed
that these two types of currents give practically the same (or at
least very similar) results in the case of the static properties
of light baryons, {\it e.g.}, magnetic moments.  This result is
easily understood because the vector and tensor currents of the
baryon octet become degenerate in the nonrelativistic limit. Also,
the magnetic moments of light baryons are dominated by the
nonrelativistic contributions, with relativistic corrections being
of higher order and small.  This explains why the simple
nonrelativistic quark approaches work so well in the description
of the magnetic moments of light baryons. Therefore, in order to
distinguish between the two types of currents of the baryon octet
we need to examine quantities which are dominated by relativistic
effects.  Two such quantities are the well known ratios $E2/M1$
and $C2/M1$ of the multipole amplitudes characterizing the $N \to
\Delta \gamma$ transition. Here we find that the sole use of
vector and tensor currents gives {\it opposite} results for the
signs of these ratios. In particular, the use of the pure vector
current for the proton gives reasonable results for $E2/M1$ and
$C2/M1$ both with a {\it correct} (negative) sign, while the use
of the pure tensor current yields ratios with {\it wrong}
(positive) sign.  Therefore, the study of the ratios $E2/M1$ and
$C2/M1$ allows one to select the appropriate current for the
description of the bound-state structure of the baryon octet
(nucleons and hyperons). It is interesting to note that in the QCD
sum rule method~\cite{Ioffe:1982ce} dealing with current quarks
the vector current structure is also preferred. This choice
originally gave an explanation of the nucleon mass, while the use
of the tensor current yields a suppression of the nucleon mass due
to the ``bad'' chiral properties of this type of the three-quark
current. We would like to stress, however, that this preference of
the vector current for the description of the baryon octet in our
approach and in QCD sum rules is apparently just coincidental
because here we are dealing with constituent quarks instead of
current quarks. Later on we will discuss why the tensor current
fails for the ratios $E2/M1$ and $C2/M1$.

First we give a summary of the results obtained. In Table 1 we
present our results for the magnetic moments of nucleons, hyperons
and nondiagonal transitions $\Sigma^0 \to \Lambda \gamma$ and $N
\to \Delta \gamma$ using the canonical set (Set
I)~\cite{Ivanov:1996pz}-\cite{Faessler:2006ft} of parameters for
the constituent quark masses $m_u = m_d = 420$ MeV and $m_s = 570$
MeV and the dimensional parameter $\Lambda_B = 1.25$ GeV
characterizing the distribution of quarks in light baryons. 
For the octet baryon states we use the vector current. Another two
solutions (Set II and Set III) corresponding to fixed values of
the constituent quark masses, but with $\Lambda_B = 0.8$ GeV and
$\Lambda_B = 0.75$ GeV, are also presented in Table 1. The reason
for decreasing the value of the dimensional parameter $\Lambda_B$
from 1.25 GeV to 0.75 GeV will be discussed below.

In Table 2 we show for comparison
the results for the magnetic moments of baryons using a pure vector
or tensor current for the octet states. The model parameters are fixed
as $m_u = m_d = 420$ MeV, $m_s = 570$ MeV and $\Lambda_B = 0.8$~GeV.
Here, for convenience, we restrict to the bare results (contributions
of valence quarks only).

In Tables 3, 4, and 5 we give our results for observables of the
$N \to \Delta \gamma$ transition such as the ratios EMR and CMR
(at zero recoil and finite $Q^2 = 0.06$ GeV$^2$), the helicity
amplitudes, the form factors $G_{E2}$, $G_{M1}$ and $G_{C2}$ at
zero recoil, the dipole $\mu_{N\Delta}$ and quadrupole
$Q_{N\Delta}$ moments, and the decay width.  Again we give results
for the three sets of parameters: Set I ($\Lambda_B = 1.25$ GeV),
Set II ($\Lambda_B = 0.8$ GeV) and Set III ($\Lambda_B = 0.75$
GeV) while using the vector current for the octet baryons. In all
cases, that is for Tables 1,3,4 and 5, we show the contributions
both of the valence quarks $(3q)$ and of the meson cloud. Note
that the mesonic cloud contribution to the magnetic moments has
been calculated~\cite{Faessler:2005gd} with the use of the chiral
Lagrangian~(\ref{L_exp}) and is expressed in terms of the
following parameters: the constituent quark mass $m$, the axial
quark charge $g$ in the chiral limit, the low-energy coupling
constants $c_2$, $c_6$, $e_7$, and $e_8$. As in
Ref.~\cite{Faessler:2005gd} the parameters $m$ and $g$ are fixed
input parameters. The value of the parameter $c_2$ has been
deduced from the analysis of the nucleon mass, meson-nucleon
sigma-terms and the $q^2$-dependence of electromagnetic nucleon
form factors. The remaining parameters $c_6$, $e_7$ and $e_8$,
controlling the size of the meson cloud contribution to the
magnetic moments, are directly fitted to reproduce the
experimental values of $\mu_p$, $\mu_n$ and $\mu_\Lambda$.

The set of values for $c_6$, $e_7$ and $e_8$ used here differs from the
ones of Ref.~\cite{Faessler:2005gd}, where instead the valence quark
contributions have been fixed using gauge, isospin and chiral symetry
constraints (as also discussed in Sec.IIc). In addition, in
Ref.~\cite{Faessler:2005gd} we additionally studied the implementation
of the naive SU(6) valence quark model and corrections, expressed for
example in the valence quark form factors of hyperons.

To state it clearly, in the present context we again use and test
important symmetry constraints already derived in
Ref.~\cite{Faessler:2005gd} and discussed in Sec.IIc.
But now we calculate the contribution of the valence quarks using the
relativistic quark model~\cite{Ivanov:1996pz}-\cite{Faessler:2006ft}.
Most of the constraints (like gauge and isospin invariance constraints)
are satisfied automatically. Fulfillment of the so-called
chiral constraints
\eq\label{chiral}
& &1 + F_2^{pu}(0) - F_2^{pd}(0) = G_2^{pu}(0) - G_2^{pd}(0) =
\biggl(\frac{g_A}{g}\biggr)^2\, \frac{m_N}{\bar m}\,, \\
& &1 + F_2^{nd}(0) - F_2^{nu}(0) = G_2^{nd}(0) - G_2^{nu}(0) =
\biggl(\frac{g_A}{g}\biggr)^2 \, \frac{m_N}{\bar m}
\nonumber
\en
is nontrivial. The direct calculation of the valence quark form
factors $F_2^{Nq}(0)$ and $G_2^{Nq}(0)$ shows a slight violation
of these identities~(\ref{chiral}). In particular,
for the Set I, II and III we get, respectively:
\eq\label{checkI}
& &1 + F_2^{pu}(0) - F_2^{pd}(0) \equiv 1 + F_2^{nd}(0) - F_2^{nu}(0)
= 4.06 \,, \\
& &G_2^{pu}(0) - G_2^{pd}(0) \equiv G_2^{nd}(0) - G_2^{nu}(0)
= 3.51 \,,
\en

\eq\label{checkII}
& &1 + F_2^{pu}(0) - F_2^{pd}(0) \equiv 1 + F_2^{nd}(0) - F_2^{nu}(0)
= 4.25 \,, \\
& &G_2^{pu}(0) - G_2^{pd}(0) \equiv G_2^{nd}(0) - G_2^{nu}(0)
= 3.62 \,,
\en
and
\eq\label{checkIII}
& &1 + F_2^{pu}(0) - F_2^{pd}(0) \equiv 1 + F_2^{nd}(0) - F_2^{nu}(0)
= 4.26 \,, \\
& &G_2^{pu}(0) - G_2^{pd}(0) \equiv G_2^{nd}(0) - G_2^{nu}(0)
= 3.63 \,.
\en
Due to the importance of satisfying these chiral constraints (they
follow from the infrared singularities of the nucleon form
factors and are model-independent identities), it is necessary to
modify the vector $\bar q \gamma_\mu q$ or the tensor 
$\bar q \sigma_{\mu\nu} q$ currents used in the evaluation of the 
$F_{1(2)}^{Nq}$ or $G_2^{Nq}$ form factors. In the following we argue 
that a modification of the $G_2^{Nq}$ form factors only leads to  
a fulfullment of these constraints. 
From Eq.~(\ref{mu_bare}) one can see, 
that a modification of the $F_{1}^{Nq}$ and $F_{2}^{Nq}$ form 
factors leads to a modification of the bare contribution 
of the baryon magnetic moment since the quantity $f_D^q(0) \equiv e_q$ 
is fixed by charge conservation. Therefore, a modification of 
the $F_{1}^{Nq}$ and $F_{2}^{Nq}$ form factors is not possible since 
it would lead to a modification of physical quantities ({\it e.g.} 
the bare baryon magnetic moments). This is not the case for the 
$G_2^{Nq}$ form factors. We can modify the $G_2^{Nq}$ form factors 
($G_2^{Nq} \to \tilde G_2^{Nq}$) to 
guarantee the fulfillment of the chiral constraints~(\ref{chiral}). 
Then we need to modify the form factors $f_P^q(0) \to \tilde f_P^q(0)$ 
to guarantee invariance of the meson-cloud contributions  
$\mu_N^{\rm cloud}$ (see Eq.~(\ref{mu_cloud}):  
\eq 
\mu_N^{\rm cloud} \, \equiv \, \sum_{q=u,d} f_P^q(0)
G_2^{Nq}(0) \, \equiv \, \sum_{q=u,d} \tilde f_P^q(0) \tilde
G_2^{Nq}(0) \,. 
\en 
On the other hand the modication $f_P^q(0) \to \tilde f_P^q(0)$  
can be achieved by a redefinition of the low-energy constants 
$c_6, e_7$ and $e_8$, which are free parameters in the chiral 
Lagrangian~(\ref{L_exp}).

The modification of the $G_2^{Nq}$ form factors is achieved
by appending the so-called ``chiral'' counterterm constructed with the
use of nucleon fields. 
In particular, the tensor currents originally 
constructed in terms of quark fields and used for the calculation
of the matrix element
$\la B(p^\prime)| \, j_{\mu\nu, q}^{\rm bare}(0) \, |B(p) \ra$
should be modified by adding a term containing nucleon fields,
viz.:
\eq\label{chiral_CT}
\bar q(x) \sigma_{\mu\nu} q(x) &\to&
\bar q(x) \sigma_{\mu\nu} q(x) \, + \,
\frac{\bar m}{m_N} \, \bar N(x) \sigma_{\mu\nu} \,
\delta G^{Nq} N(x) \,,
\en
where $\delta G^{Nu} = {\rm diag}\{\delta G^{pu}, \delta G^{nu}\}$
and $\delta G^{Nd} = {\rm diag}\{\delta G^{pd}, \delta G^{nd}\}$
are the diagonal $2 \times 2$ flavor matrices and $q = u$ or~$d$.

These matrix elements are fixed to enhance the magnitudes of the
form factors $G_2^{Nq}(0)$ and to satisfy the chiral
constraints~(\ref{chiral_constr2}). The idea is to increase the
combinations $G_2^{pu}(0) - G_2^{pd}(0)$ and
$G_2^{nd}(0) - G_2^{nu}(0)$ from
3.51 (Set I), 3.62 (Set II) or 3.63 (Set III) to 
4.06 (Set I), 4.25 (Set II) or 4.26 (Set III).
Note that such modifications alter the
normalization of the form factors $G_2^{Nq}$ without additional
change of these form factors at finite values of $q^2$.
To fulfill the chiral constraints~(\ref{chiral}), we fix the
constants $\delta G^{Nq}$ in~(\ref{chiral_CT}) as:

\vspace*{.1cm}

Set I

\eq
\delta G^{pu} \equiv \delta G^{nd} \, = \, 
- 4 \delta G^{pd} \equiv - 4 \delta G^{nu} = 0.440 \,,
\en

Set II

\eq
\delta G^{pu} \equiv \delta G^{nd} \, = \,  
- 4 \delta G^{pd} \equiv - 4 \delta G^{nu} = 0.504 \,,
\en

Set III

\eq \delta G^{pu} \equiv \delta G^{nd} \, = \, 
- 4 \delta G^{pd} \equiv - 4 \delta G^{nu} = 0.504 \,. 
\en 
After introducing the counterterm~(\ref{chiral_CT}) the form
factors $G_2^{Nq}$ are modified as 
\eq G_2^{Nq}(0) \to \tilde
G_2^{Nq}(0) = G_2^{Nq}(0) +  \delta G_2^{Nq}(0) \,. 
\en 
As we stressed before we need to modify the form factors 
$f_P^q(0) \to \tilde f_P^q(0)$ to guarantee invariance of 
the meson-cloud contributions $\mu_N^{\rm cloud}$. 
As stated above, the result for $\tilde
f_P^q(0)$ has no physical meaning: we merely need to redefine the
low-energy couplings parametrizing this quantity. 

Finally, to guarantee the invariance of the meson-cloud contributions
to the magnetic moments of other baryons (including
$N \to \Delta \gamma$ transition) we need to introduce the
``chiral'' counterterms by extending the tensor quark operator
\eq\label{chiral_CT2}
\bar q(x) \sigma_{\mu\nu} q(x) &\to&
\bar q(x) \sigma_{\mu\nu} q(x) \, + \,
\frac{\bar m}{m_B} \, \sum\limits_{B}
\bar B(x) \sigma_{\mu\nu} \delta G^{Bq} B(x) + \cdots  \,,
\en
where $\delta G^{Bq}$ is fixed from the condition
\eq
\mu_B^{\rm cloud} \, \equiv \, \sum_{q=u,d,s} f_P^q(0) G_2^{Bq}(0)
\, \equiv \, \sum_{q=u,d,s} \tilde f_P^q(0) \tilde G_2^{Bq}(0) \,.
\en
Note that we do not modify the form factors associated with
the strange quark, that is $f_P^s \equiv \tilde f_P^s$ and
$G_2^{Bs} \equiv \tilde G_2^{Bs}$, because we do not have special
constraints on the strange quark contributions.
In Eq.~(\ref{chiral_CT2}) we display for transparency only the diagonal
operators in terms of baryon fields.
The nondiagonal terms relevant for $\Sigma^0 \to \Lambda \gamma$
and $N \to \Delta \gamma$ transitions are omitted (symbol $\cdots$)
and can be derived in analogy.

Now we discuss, how the parameters intering in the calculation 
of meson-cloud contributions are determined. 
With the use of the chiral constraint~(\ref{chiral}),
the physical mass of the nucleon $m_N = m_p = 938.27$ MeV and its
axial charge $g_A = 1.267$~\cite{Eidelman:2004wy} we fixed the 
quark axial charge as: 
$g = 0.94$ (Set I), $g = 0.92$ (Set II), $g = 0.92$ (Set III).
For the parameter $c_2$ we use the value fixed in
Ref.~\cite{Faessler:2005gd}: $c_2 = 2.502$ GeV$^{-1}$. 
The parameters $c_6$, $e_7$ and $e_8$ are fixed as: 

\vspace*{.1cm}

Set I

\eq\label{SetI} 
\tilde c_6 = 0.163                   \,, \,\,\,
\bar e_7 = - 0.426  \ {\rm GeV}^{-3} \,, \,\,\,
\bar e_8 = - 0.097  \ {\rm GeV}^{-3} \,.
\en

Set II

\eq\label{SetII} 
\tilde c_6 = 0.067                   \,, \,\,\,
\bar e_7 = - 0.318  \ {\rm GeV}^{-3} \,, \,\,\,
\bar e_8 = - 0.076  \ {\rm GeV}^{-3} \,.
\en

Set III
\eq\label{SetIII} 
\tilde c_6 = 0.067                   \,, \,\,\,
\bar e_7 = - 0.314  \ {\rm GeV}^{-3} \,, \,\,\,
\bar e_8 = - 0.082 \ {\rm GeV}^{-3} \,.
\en

In the calculation of the $q^2$-dependence of the meson-cloud
contribution following constants in the chiral Lagrangian~(\ref{L_qU})
enter: $c_4$, $d_{10}$ and $e_{10}$. Here for $c_4$, $d_{10}$ and
$e_{10}$ we use the values fixed previously~\cite{Faessler:2005gd}:
\eq\label{c4d10e10}
c_4 = 1.693  \ {\rm GeV}^{-1} \,, \,\,\,
\bar d_{10} = 1.110  \ {\rm GeV}^{-2} \,, \,\,\,
\bar e_{10} = 0.039  \ {\rm GeV}^{-3} \,.
\en
In Eqs.~(\ref{SetI})-(\ref{SetIII}) the constants
$\bar d_{10}$, $\bar e_7$ and $\bar e_8$ and $\bar e_{10}$
refer to the renormalized coupling constants
(see details in Ref.~\cite{Faessler:2005gd}) and
$\tilde c_6 = c_6 - 16 m (2 \hat m + \hat m_s)$.

Now we return to discuss our results. As evident from Table 1 the
magnetic moments of the baryon octet can be described with good
accuracy for different values of the parameter $\Lambda_B$. There
is only a weak dependence on the variation of this parameter from
0.8 GeV to 1.25 GeV. However, the dipole moment $\mu_{N\Delta}$ of
the $N \to \Delta \gamma$ transition is quite sensitive to the
variation of $\Lambda_B$. The reason for this stronger dependence
is that $\mu_{N\Delta}$ is proportional to the combination of the
form factors $b_3$ and $b_2$: \eq \mu_{N\Delta} \, \sim  \, b_3(0)
\frac{m_+ (3 m_\Delta + m_N)}{m_\Delta} \, + \, b_2(0) m_+ m_- \,.
\en The main contribution to $\mu_{N\Delta}$ comes from the
$b_3(0)$ form factor which has dimension $1/M$. If we restrict our
attention to the leading contribution to $\mu_{N\Delta}$ coming
from the $3q$ core then we immediately realize that $b_3$ scales
as $1/\Lambda_B$. Hence we need to decrease the parameter
$\Lambda_B$ to get a reasonable description of $\mu_{N\Delta}$.
With $\Lambda_B \simeq 0.75$ GeV one can fit the central value of
$\mu_{N\Delta}$ precisely.

The next point of discussion is the sign of the EMR and CMR
ratios. Again we restrict our attention to zero recoil~-- $Q^2 =
0$. The contribution of the $b_4$ form factor can essentially be
neglected in our considerations since we find $b_4(0)/b_3(0)
\simeq - 1/10$ and $b_4(0)/b_2(0) \simeq 1/5$. If we temporarily
also neglect the $b_2$ form factor in EMR and CMR then (this is a
well-known result in the
literature~\cite{Bjorken:1966ij}-\cite{Braun:2005be}) these ratios
become degenerate and equal to \eq {\rm EMR} \ \, = \, \ {\rm CMR}
\ \, = \, - \, \frac{m_-}{3 m_\Delta + m_N} \, \simeq - 6 \% \,.
\en Therefore, to reproduce the {\it experimental} results for
these quantities we require a contribution from the $b_2$ form
factor. In both cases (for a vector and a tensor current) the
value of the $b_2$ form factor is negative, but in the case of the
tensor current it is twice as large than what is required
phenomenologically. As a result, in the case of the tensor proton
current the $G_{E2}$ and $G_{C2}$ form factors actually change
sign from positive to negative, leading to positive ratios EMR and
CMR. This is {\it not} the case for the vector current for the
proton, and we therefore conclude that the vector current is
strongly preferred in the calculations of the properties of $N \to
\Delta \gamma$ transition. To further illustrate this issue, in
Table 6 we demonstrate the sensitivity of the EMR and CMR ratios
on the choice of the three-quark proton current (see discussion
below).

Again, Table 2 shows that the pure vector and
the pure tensor current used for the baryon octet give similar results
for the bare magnetic moments of light baryons for the same set of
model parameters: constituent quark masses and
dimensional parameter $\Lambda_B$. In Table 2 we restrict ourselves to the
specific choice of model parameters (Set II):
$m_u = m_d = 420$ MeV, $m_s = 570$ MeV and $\Lambda_B = 0.8$ GeV.
However, the {\it similarity} of results for the two respective octet
currents is not very sensitive to a variation of the model parameters.

In Tables 3, 4 and 5 we present the detailed results for the
properties of the $N \to \Delta \gamma$ transition for different
values of the dimensional parameter $\Lambda_B = $ 1.25, 0.8 and 0.75 GeV,
respectively. For the EMR and CMR ratios we present our predictions
at zero recoil ($Q^2 = 0$) and at the finite value $Q^2 = 0.06$
GeV$^2$ (recently the A1 Collaboration at
Mainz~\cite{Stave:2006ea} measured these quantities at this
kinematic point). Our predictions are in good
agreement with the experimental data of the LEGS Collaboration
at Brookhaven~\cite{Blanpied:2001ae} and of the GDH, A1 and A2
Collaborations at Mainz~\cite{Ahrens:2004pf,Stave:2006ea}. The quantities
which are sensitive to the choice of the dimensional parameter $\Lambda_B$
are the magnetic, electric and Coulomb form factors and related
quantities -- helicity amplitudes, dipole and quadrupole moment, decay
width. As we stressed before, the magnetic form factor $G_{M1}$ and the
dipole moment $\mu_{N\Delta}$ increase when the parameter $\Lambda_B$
decreases. Other quantities mentioned above have the same tendency.
Therefore, the best description of the data is achieved for values of
$\Lambda_B = 0.8$ or $0.75$ GeV. In Figs.4-10 we demonstrate the dependence
of the $G_{M1}$, $G_{E2}$, $G_{C2}$ form factors, the helicity amplitudes
$A_{1/2}$ and $A_{3/2}$, and the ratios EMR and CMR as functions of
$Q^2$ up to values of 0.2 GeV$^2$. Results are indicated for the parameter
Set II with $\Lambda_B = 0.8$ GeV. In the figures the solid line
corresponds to the total contribution while the dashed line marks the bare
contribution or the one of the valence quarks.

Future refinement of the present work will involve tests of the functional
form of the vertex functions entering into the strong interaction
Lagrangian~(\ref{Lagr_str}) as well as
the  form of the quark propagator modified in order to account for
confinement. More precise data will also allow to study the possible
mixture of vector and tensor currents.

In Table 6 we demonstrate the sensitivity of the EMR and CMR ratios
at $Q^2=0$ on the choice of the proton three-quark current for
typical values of the parameter $\Lambda_B = 0.75, 0.8$ and $1.25$ GeV.
The proton current is used in the form
\eq
J_p \, = \, (1- \beta) J_p^V \, + \, \beta J_p^T
\en
where
$\beta$ is a tensor-vector mixing parameter. The limiting
cases $\beta = 0$ and $\beta = 1$ correspond to the use of pure
vector and pure tensor currents, respectively. We hope that forthcoming
more precise experiments on the ratios EMR and CMR can yield a
relatively precise limit on the value of the mixing parameter $\beta$.

\section{Summary}

In this paper we have calculated the magnetic moments of light
baryons as well as the $N \to \Delta \gamma$ transition
properties using a manifestly Lorentz covariant chiral quark
approach to the study of baryons as bound states of constituent
quarks dressed by a cloud of pseudoscalar mesons. Our main results
are:

- The contribution of the meson cloud to the static properties of
light baryons is up to 20\%, which is consistent with the
perturbative nature of their contribution and, together with the
relativistic corrections, helps to explain how the ~30\% shortfall
in the SU(6) prediction is ameliorated;

- We showed that the numerical value of the dipole magnetic moment
$\mu_{N\Delta}$ is sensitive to the scale parameter $\Lambda_B$
describing the distribution of quarks in the baryon. In
particular, this quantity scales as $1/\Lambda_B$ and a reasonable
description of data is achieved at $\Lambda_B \leq 0.8$ GeV due to
the enhancement of the valence quark contribution;

- The multipole ratios EMR and CMR are sensitive to the choice of the
  proton current: vector $J_p^V$ or tensor $J_p^T$ (see Appendix~A).
  The use of a pure vector current $J_p^V$ gives a reasonable
  description of the data. The pure tensor current $J_p^T$ gives results
  for EMR and CMR with the wrong (positive) sign.  However, a small
  admixture of the tensor current is possible, and forthcoming
  experiments can give a strong restriction on the mixing parameter of
  such currents;

- We presented a detailed analysis of the light baryon observables
  all of which are in good agreement with experimental data.

\begin{acknowledgments}

This work was supported by the DFG under contracts FA67/31-1 and
GRK683. This research is also part of the EU Integrated
Infrastructure Initiative Hadronphysics project under contract
number RII3-CT-2004-506078 and President grant of Russia
"Scientific Schools"  No. 5103.2006.2. K.P. thanks the Development
and Promotion of Science and Technology Talent Project (DPST),
Thailand for financial support. BRH is supported by the US
National Science Foundation under award PHY 02-44801 and would
like to thank the T\"ubingen theory group for its hospitality.

\end{acknowledgments}

\newpage

\appendix\section{Three-quark baryon currents}

Here we specify the baryonic currents~\cite{Ioffe:1982ce,Efimov:1987na}.
The three-quark currents of the baryon octet are:

\vspace*{.2cm}

\centerline{I. Vector currents}
\eq
J_p^V &=& \varepsilon^{a_1a_2a_3} \gamma^\mu \gamma^5 d^{a_1} u^{a_2}
C \gamma_\mu u^{a_3}  \,, \nonumber\\[2mm]
J_n^V &=& - \varepsilon^{a_1a_2a_3} \gamma^\mu \gamma^5 u^{a_1} d^{a_2}
C \gamma_\mu d^{a_3}  \,, \nonumber\\[2mm]
J_{\Sigma^+}^V &=& - \varepsilon^{a_1a_2a_3}
\gamma^\mu \gamma^5 s^{a_1} u^{a_2}
C \gamma_\mu u^{a_3}  \,, \nonumber\\[2mm]
J_{\Sigma^0}^V &=& \sqrt{2} \ \varepsilon^{a_1a_2a_3}
\gamma^\mu \gamma^5 s^{a_1} u^{a_2}
C \gamma_\mu d^{a_3} \,, \\[2mm]
J_{\Sigma^-}^V &=& \varepsilon^{a_1a_2a_3}
\gamma^\mu \gamma^5 s^{a_1} d^{a_2}
C \gamma_\mu d^{a_3}  \,, \nonumber\\[2mm]
J_{\Xi^-}^V &=& \varepsilon^{a_1a_2a_3}
\gamma^\mu \gamma^5 d^{a_1} s^{a_2}
C \gamma_\mu s^{a_3}  \,, \nonumber\\[2mm]
J_{\Xi^0}^V &=& \varepsilon^{a_1a_2a_3}
\gamma^\mu \gamma^5 u^{a_1} s^{a_2} C \gamma_\mu s^{a_3}  \,,
\nonumber\\[2mm]
J_{\Lambda^0}^V &=& \sqrt{\frac{2}{3}} \ \varepsilon^{a_1a_2a_3}
\gamma^\mu \gamma^5 (u^{a_1} d^{a_2} C \gamma_\mu s^{a_3} -
d^{a_1} u^{a_2} C \gamma_\mu s^{a_3} )  \,. \nonumber
\en
\centerline{II. Tensor currents}
\eq
J_p^T &=& \varepsilon^{a_1a_2a_3} \sigma^{\mu\nu} \gamma^5 d^{a_1} u^{a_2}
C \sigma_{\mu\nu}  u^{a_3}  \,, \nonumber\\[2mm]
J_n^T &=& - \varepsilon^{a_1a_2a_3} \sigma^{\mu\nu}  \gamma^5 u^{a_1} d^{a_2}
C \sigma_{\mu\nu}  d^{a_3}  \,, \nonumber\\[2mm]
J_{\Sigma^+}^T &=& - \varepsilon^{a_1a_2a_3}
\sigma^{\mu\nu}  \gamma^5 s^{a_1} u^{a_2}
C \sigma_{\mu\nu} u^{a_3}  \,, \nonumber\\[2mm]
J_{\Sigma^0}^T &=& \sqrt{2} \ \varepsilon^{a_1a_2a_3}
\sigma^{\mu\nu}  \gamma^5 s^{a_1} u^{a_2}
C \sigma_{\mu\nu}  d^{a_3} \,, \\[2mm]
J_{\Sigma^-}^T &=& \varepsilon^{a_1a_2a_3}
\sigma^{\mu\nu} \gamma^5 s^{a_1} d^{a_2}
C \sigma_{\mu\nu} d^{a_3}  \,, \nonumber\\[2mm]
J_{\Xi^-}^T &=& \varepsilon^{a_1a_2a_3}
\sigma^{\mu\nu}  \gamma^5 d^{a_1} s^{a_2}
C \sigma_{\mu\nu}  s^{a_3}  \,, \nonumber\\[2mm]
J_{\Xi^0}^T &=& \varepsilon^{a_1a_2a_3}
\sigma^{\mu\nu} \gamma^5 u^{a_1} s^{a_2} C \sigma_{\mu\nu}  s^{a_3}  \,,
\nonumber\\[2mm]
J_{\Lambda^0}^T &=& \sqrt{\frac{2}{3}} \ \varepsilon^{a_1a_2a_3}
\sigma^{\mu\nu}  \gamma^5 (u^{a_1} d^{a_2} C \sigma_{\mu\nu}  s^{a_3} -
d^{a_1} u^{a_2} C \sigma_{\mu\nu} s^{a_3} )  \,. \nonumber
\en
The three-quark (vector) currents of the $\Delta$-isobar are:
\eq\label{currents_D1}
J_{\Delta^{++}}^\mu &=&
\varepsilon^{a_1a_2a_3} u^{a_1} u^{a_2}
C \gamma^\mu u^{a_3}  \,, \nonumber\\[2mm]
J_{\Delta^{+}}^\mu &=& \frac{1}{\sqrt{3}}
\varepsilon^{a_1a_2a_3} (d^{a_1} u^{a_2}
C \gamma^\mu u^{a_3}  +
2 u^{a_1} u^{a_2} C \gamma^\mu d^{a_3})
\,, \nonumber\\[2mm]
J_{\Delta^{0}}^\mu &=& \frac{1}{\sqrt{3}}
\varepsilon^{a_1a_2a_3} (u^{a_1} d^{a_2}
C \gamma^\mu d^{a_3}  +
2 d^{a_1} d^{a_2} C \gamma^\mu u^{a_3}) \,, \nonumber\\[2mm]
J_{\Delta^{-}}^\mu &=&
\varepsilon^{a_1a_2a_3} d^{a_1} d^{a_2}
C \gamma^\mu d^{a_3}  \,.
\en
In the case of the $\Delta^+$ and $\Delta^0$ states it is also useful
to proceed with the currents which contain two identical quarks
(two ``up'' or two ``down'' quarks) contracted together as a diquark
subsystem:

\eq\label{currents_D2}
J_{\Delta^{+}}^\mu &=& \frac{1}{\sqrt{3}}
\varepsilon^{a_1a_2a_3}
(2 d^{a_1} u^{a_2} C \gamma^\mu u^{a_3}
- i \gamma_\nu d^{a_1} u^{a_2} C \sigma^{\mu\nu} u^{a_3})\,,
\nonumber \\
J_{\Delta^{0}}^\mu &=& \frac{1}{\sqrt{3}}
\varepsilon^{a_1a_2a_3}
(2 u^{a_1} d^{a_2} C \gamma^\mu d^{a_3}
- i \gamma_\nu u^{a_1} d^{a_2} C \sigma^{\mu\nu} d^{a_3})\,.
\en
Eqs.~(\ref{currents_D2}) are derived from Eqs.~(\ref{currents_D1})
using Fierz transformations.

\section{Sets of the relativistic form factors
for the $N\to\Delta\gamma$ transition}

In the literature one can find several equivalent decompositions
of the vertex function $\Lambda_{\mu\nu}(p,p^\prime)$ describing the
$N \to \Delta \gamma$ transition~\cite{Bjorken:1966ij}-\cite{Braun:2005be}

\eq
\Lambda_{\mu\nu}^{(1)}(p,p^{\prime}) &=&
[ ( \gamma_\mu q_\nu  - g_{\mu\nu} \not\! q ) G_1(q^2) +
( p^\prime_\mu q_\nu - g_{\mu\nu} p^\prime q ) G_2(q^2) +
( q_\mu q_\nu - g_{\mu\nu} q^2 )  G_3(q^2) ] \gamma^5  \,, \nonumber\\
\Lambda_{\mu\nu}^{(2)}(p,p^{\prime}) &=& 
[ (g_{\mu\nu} q^2 + q_\mu p_\nu) a_1(q^2) +
a_2 (g_{\mu\nu} m_+ m_- + P_\mu p_\nu) a_2(q^2) +
(g_{\mu\nu} m_+ + \gamma_\mu p_\nu) a_3(q^2) ] \gamma^5  \,, \\
\Lambda_{\mu\nu}^{(3)}(p,p^{\prime}) &=& -  \frac{1}{2m_N}
[ (g_{\mu\nu} \not\! q - \gamma_\mu q_\nu) c_1(q^2) +
(g_{\mu\nu} q^2 - q_\mu q_\nu) \frac{c_2(q^2)}{2m_N} +
(g_{\mu\nu} pq - p_\mu q_\nu)  \frac{c_3(q^2)}{2m_N} + 
g_{\mu\nu} c_4(q^2) ] \gamma^5 \,.
\nonumber
\en
where $P = p + p^\prime$ and $m_\pm = m_\Delta \pm m_N$.
The sets of the relativistic form factors $G_i$, $a_i$,
$b_i$ [see Eq.~(\ref{set_b})] and $c_i$ are related to each other as:
\eq
& &G_1 = - a_3 = b_3 = \frac{c_1}{2m_N} \,, \nonumber\\
& &G_2 = - 2a_2 = b_2 = \frac{c_3}{4m_N^2} \,, \nonumber\\
& &G_3 = - a_1 + a_2 = - b_2 + b_4 = \frac{c_2 - c_3}{4m_N^2} \,,
\nonumber\\
& &c_4 \equiv 0 \,.
\en

\newpage

\vspace*{2cm}

\noindent

\begin{center}

\def\arraystretch{1.5}
{\bf Table 1.} Magnetic moments of light baryons (in units of the
nuclear magneton $\mu_N$). \\ Results are calculated for the case of
a purely vector current.

\vspace*{.3cm}

\begin{tabular}{|c||c|c|c||c|c|c||c|c|c||c|}
\hline
& \multicolumn{3}{c||}{Set I ($\Lambda_B = 1.25$ GeV)}
& \multicolumn{3}{c||}{Set II ($\Lambda_B = 0.8$ GeV)}
& \multicolumn{3}{c||}{Set III ($\Lambda_B = 0.75$ GeV)} &\\
\cline{2-10}
& Bare & Meson & Total & Bare & Meson & Total & Bare & Meson & Total
& Experiment~\cite{Eidelman:2004wy,Blanpied:2001ae}\\
& (3q) & cloud &       & (3q) & cloud &       & (3q) & cloud &       & \\
\hline\hline
$\mu_p$ &\, 2.530 \,&\, 0.263 \,&\, 2.793 \,&\, 2.614 \,&\, 0.179 \,&\,
2.793 \,&\, 2.621 \,&\, 0.172 \,
&\, 2.793 \,&\, 2.793 \\
$\mu_n$ & -1.530 & -0.383 & -1.913 & -1.634 & -0.279 & -1.913
& -1.643 & -0.270 & -1.913 & -1.913 \\
$\mu_{\Lambda}$ & -0.575 & -0.038 & -0.613 & -0.579 & -0.034 & -0.613
& -0.578 & -0.035 & -0.613 &-0.613 $\pm$ 0.004 \\
$\mu_{\Sigma^+}$ & 2.336 & 0.196 & 2.532 & 2.423 & 0.148 & 2.571
& 2.430 & 0.130 & 2.560 & 2.458 $\pm$ 0.010\\
$\mu_{\Sigma^-}$ & -0.942 & -0.327 & -1.269 & -0.960 & -0.223 & -1.183
& -0.962 & -0.235 & -1.197 & -1.160 $\pm$ 0.025\\
$\mu_{\Xi^0}$ & -1.240 & -0.096 & -1.336 & -1.303 & -0.082 & -1.385
& -1.310 & -0.076 & -1.386 &-1.250 $\pm$ 0.014 \\
$\mu_{\Xi^-}$ & -0.599 & 0.033 & -0.566 & -0.567 & 0.012 & -0.555
& -0.562 & 0.014 & -0.548 & -0.6507 $\pm$ 0.003 \\
$|\mu_{\Sigma^0 \Lambda}|$ & 1.273 & 0.293 & 1.566 & 1.372 & 0.245 & 1.617
& 1.385 & 0.222 & 1.607 & 1.61 $\pm$ 0.08 \\
$\mu_{N \Delta}$ & 2.357 & 0.439 & 2.796
& 2.984 & 0.354 & 3.338 & 3.102 & 0.356 & 3.458
& 3.642 $\pm$ 0.019 $\pm$ 0.085\\
\hline
\end{tabular}
\end{center}

\vspace*{2cm}

\noindent

\begin{center}

\def\arraystretch{1.5}
{\bf Table 2.} Sensitivity of the bare contributions to
the light baryon magnetic moments on \\
the choice of the octet baryon $3q$-current (in units of the
nuclear magneton $\mu_N$). \\ The scale parameter is chosen to be
$\Lambda_B=0.8$ GeV.

\vspace*{.3cm}

\begin{tabular}{|c|c|c|c|}
\hline
& Vector current & Tensor current
& Experiment~\cite{Eidelman:2004wy,Blanpied:2001ae}\\
\hline\hline
$\mu_p$ &\, 2.614 \,&\, 2.804 \,&\, 2.793 \\
$\mu_n$ & -1.634    & -1.814 & -1.913 \\
$\mu_{\Lambda}$     & -0.579 & -0.594 & -0.613 $\pm$ 0.004 \\
$\mu_{\Sigma^+}$    & 2.423  & 2.509 & 2.458 $\pm$ 0.010\\
$\mu_{\Sigma^-}$    & -0.960 & -0.973 & -1.160 $\pm$ 0.025\\
$\mu_{\Xi^0}$       & -1.303 & -1.385 & -1.250 $\pm$ 0.014 \\
$\mu_{\Xi^-}$       & -0.567 & -0.560 & -0.6507 $\pm$ 0.003 \\
$|\mu_{\Sigma^0 \Lambda}|$   & 1.372  & 1.398 & 1.61 $\pm$ 0.08 \\
$\mu_{N \Delta}$    & 2.984  & 2.740  & 3.642 $\pm$ 0.019 $\pm$ 0.085\\
\hline
\end{tabular}
\end{center}

\newpage

\noindent

\begin{center}
\def\arraystretch{1.5}
{\bf Table 3.} Results for the
N $\to \Delta\gamma$ transition (Set I: $\Lambda_B = 1.25$ GeV)
\end{center}

\begin{center}
\begin{tabular}{|c|c|c|c|c|}
\hline
& Bare (3q) & Meson cloud & \,\,Total \,\,&
Experiment~\cite{Eidelman:2004wy,Blanpied:2001ae,Stave:2006ea}\\
\hline\hline
\hspace*{-1.4cm}EMR (\%) at $Q^2=0$
& -3.22 & 0.29 & -2.93 & -2.5 $\pm$ 0.5;
-3.07 $\pm$ 0.26 $\pm$ 0.24 \\
EMR (\%) at $Q^2=0.06$ GeV$^2$ & -3.14 & 0.42 & -2.72
& -2.28 $\pm$ 0.29 $\pm$ 0.20\\
\hspace*{-1.4cm}CMR (\%) at $Q^2=0$
& -3.69 & 0.34 & -3.35 & \\
CMR (\%) at $Q^2=0.06$ GeV$^2$ & -4.75 & 0.44 & -4.31
& -4.81 $\pm$ 0.27 $\pm$ 0.26\\
$A_{1/2}(0)$ in $10^{-3}\ $ GeV$^{-1/2}$
& -87.4 & -11.8 & -99.2 & -135 $\pm$ 6\\
$A_{3/2}(0)$ in $10^{-3}\ $ GeV$^{-1/2}$
& -173.0 & -20.9 & -193.9 & -250 $\pm$ 8\\
$G_{E2}(0)$ & 0.093 & 0.002 & 0.095    & 0.137 $\pm$ 0.012 $\pm$ 0.043\\
$G_{M1}(0)$ & 2.887 & 0.359 & 3.246 & 4.460 $\pm$ 0.023 $\pm$ 0.104\\
$G_{C2}(0)$ & 0.107 & 0.008 & 0.115 & \\
$Q_{N\Delta} \ $( fm$^2 )$&  -0.073 & -0.001 & -0.074 &
-0.108 $\pm$ 0.009 $\pm$ 0.034\\
$\mu_{N\Delta}$ &  2.357 & 0.439 & 2.796
& 3.642 $\pm$ 0.019 $\pm$ 0.085 \\
$\Gamma_{\Delta \to \gamma}$ (MeV) & 0.30 & 0.09 & 0.39
& 0.58 - 0.67 \\
\hline
\end{tabular}
\end{center}

\vspace*{.75cm}

\noindent

\begin{center}
\def\arraystretch{1.5}
{\bf Table 4.} Results for the
N $\to \Delta\gamma$ transition (Set II: $\Lambda_B = 0.8$ GeV)
\end{center}

\begin{center}
\begin{tabular}{|c|c|c|c|c|}
\hline
 & Bare (3q) & Meson cloud & \,\,Total \,\, &
Experiment~\cite{Eidelman:2004wy,Blanpied:2001ae,Stave:2006ea}\\
\hline\hline
\hspace*{-1.4cm}EMR (\%) at $Q^2=0$
& -3.41 & 0.31 & -3.10 & -2.5 $\pm$ 0.5;
-3.07 $\pm$ 0.26 $\pm$ 0.24 \\
EMR (\%) at $Q^2=0.06$ GeV$^2$ & -3.34 & 0.33 & -3.01
& -2.28 $\pm$ 0.29 $\pm$ 0.20\\
\hspace*{-1.4cm}CMR (\%) at $Q^2=0$
& -3.95 & 0.26 & -3.69 & \\
CMR (\%) at $Q^2=0.06$ GeV$^2$ & -5.13 & 0.35 & -4.78
& -4.81 $\pm$ 0.27 $\pm$ 0.26\\
$A_{1/2}(0)$ in $(10^{-3}\ $ GeV$^{-1/2})$
& -110.0 & -14.3 & -124.3 & -135 $\pm$ 6\\
$A_{3/2}(0)$ in $(10^{-3}\ $ GeV$^{-1/2})$
& -219.4 & -25.3 & -244.7 & -250 $\pm$ 8\\
$G_{E2}(0)$ & 0.125 &  0.002 & 0.127    & 0.137 $\pm$ 0.012 $\pm$ 0.043\\
$G_{M1}(0)$ & 3.655 & 0.434 & 4.089 & 4.460 $\pm$ 0.023 $\pm$ 0.104\\
$G_{C2}(0)$ & 0.144 & 0.007 & 0.151 & \\
$Q_{N\Delta} \ $( fm$^2 )$&  -0.098 & -0.001 & -0.099 &
-0.108 $\pm$ 0.009 $\pm$ 0.034\\
$\mu_{N\Delta}$ &  2.984 & 0.354 & 3.338
& 3.642 $\pm$ 0.019 $\pm$ 0.085 \\
$\Gamma_{\Delta \to N \gamma}$ (MeV) & 0.49  & 0.12 & 0.61
& 0.58 - 0.67 \\
\hline
\end{tabular}
\end{center}

\vspace*{.75cm}

\noindent 

\begin{center}
\def\arraystretch{1.5}
{\bf Table 5.} Results for the
N $\to \Delta\gamma$ transition (Set III: $\Lambda_B = 0.75$ GeV)
\end{center}

\begin{center}
\begin{tabular}{|c|c|c|c|c|}
\hline
& Bare (3q) & Meson cloud & \,\,Total \,\,
& Experiment~\cite{Eidelman:2004wy,Blanpied:2001ae,Stave:2006ea}\\
\hline\hline
\hspace*{-1.4cm}EMR (\%) at $Q^2=0$
& -3.43 & 0.30 & -3.13 & -2.5 $\pm$ 0.5;
-3.07 $\pm$ 0.26 $\pm$ 0.24 \\
EMR (\%) at $Q^2=0.06$ GeV$^2$ & -3.35 & 0.30 & -3.05
& -2.28 $\pm$ 0.29 $\pm$ 0.20\\
\hspace*{-1.4cm}CMR (\%) at $Q^2=0$
& -3.98 & 0.25 & -3.73 & \\
CMR (\%) at $Q^2=0.06$ GeV$^2$ & -5.17 & 0.33 & -4.84
& -4.81 $\pm$ 0.27 $\pm$ 0.26\\
$A_{1/2}(0)$ in $(10^{-3}\ $ GeV$^{-1/2})$
& -114.3 & -14.3 & -128.6 & -135 $\pm$ 6\\
$A_{3/2}(0)$ in $(10^{-3}\ $ GeV$^{-1/2})$
& -228.1 & -25.4 & -253.5 & -250 $\pm$ 8\\
$G_{E2}(0)$ & 0.130 & 0.002 & 0.132 & 0.137 $\pm$ 0.012 $\pm$ 0.043\\
$G_{M1}(0)$ & 3.800 & 0.435 & 4.235 & 4.460 $\pm$ 0.023 $\pm$ 0.104\\
$G_{C2}(0)$ & 0.151 & 0.007 & 0.158 & \\
$Q_{N\Delta} \ $( fm$^2 )$&  -0.102 & -0.002 & -0.104 &
-0.108 $\pm$ 0.009 $\pm$ 0.034\\
$\mu_{N\Delta}$ & 3.102 & 0.356 & 3.458
& 3.642 $\pm$ 0.019 $\pm$ 0.085 \\
$\Gamma_{\Delta \to N \gamma}$ (MeV) &  0.53 & 0.13 & 0.66
& 0.58 - 0.67 \\
\hline
\end{tabular}
\end{center}

\newpage

\vspace*{2cm}

\noindent

\begin{center}
\def\arraystretch{1.5}
{\bf Table 6.} Sensitivity of the EMR and CMR ratios
to the choice of the proton $3q$-current

\vspace*{.3cm}

\begin{tabular}{|c||c|c|c|c|c|c|c|c|c|c|c|c|c|c||}
\hline
& \multicolumn{14}{c||}{Mixing parameter $\beta$} \\
\cline{2-15}
& 0 & 0.025 & 0.05 & 0.075 & 0.1 & 0.15 & 0.2 & 0.25 & 0.3
& 0.35 & 0.4 & 0.5 & 0.75 & 1  \\
\hline
Set I ($\Lambda_B = 1.25$ GeV)  & & & & & & & & & & & & & & \\
EMR (\%)& -2.93 & -2.54 & -2.28 & -2.04 & -1.80
        & -1.35 & -1.08 & -0.55 & -0.19 & 0.15
        & 0.47  & 1.05  & 2.29  & 3.19 \\
CMR (\%)& -3.35 & -3.03 & -2.72 & -2.42 & -2.13
        & -1.59 & -1.08 & -0.61 & -0.17 & 0.25
        & 0.63  & 1.35  & 2.81  & 3.95 \\
\hline
Set II ($\Lambda_B = 0.8$ GeV)  & & & & & & & & & & & & & & \\
EMR (\%)& -3.10 & -2.83 & -2.56 & -2.30 & -2.06
        & -1.60 & -1.17 & -0.77 & -0.40 & -0.05
        & 0.28  & 0.87  & 2.09  & 3.03 \\
CMR (\%)& -3.69 & -3.35 & -3.03 & -2.64 & -2.35
        & -1.80 & -1.29 & -0.81 & -0.37 & 0.04
        & 0.43  & 1.14  & 2.58  & 3.69 \\
\hline
Set III ($\Lambda_B = 0.75$ GeV) & & & & & & & & & & & & & & \\
EMR (\%)& -3.13 & -2.84  & -2.58 & -2.33 & -2.07
        & -1.62 & -1.19  & -0.79 & -0.41 & -0.07
        & 0.26  & 0.85   & 2.10  & 3.00 \\
CMR (\%)& -3.73 & -3.39  & -3.06 & -2.75 & -2.44
        & -1.87 & -1.34  & -0.85 & -0.40 & 0.03
        & 0.43  & 1.16   & 2.65  & 3.80 \\
\hline
\end{tabular}
\end{center}

\newpage

\begin{center}
\epsfig{figure=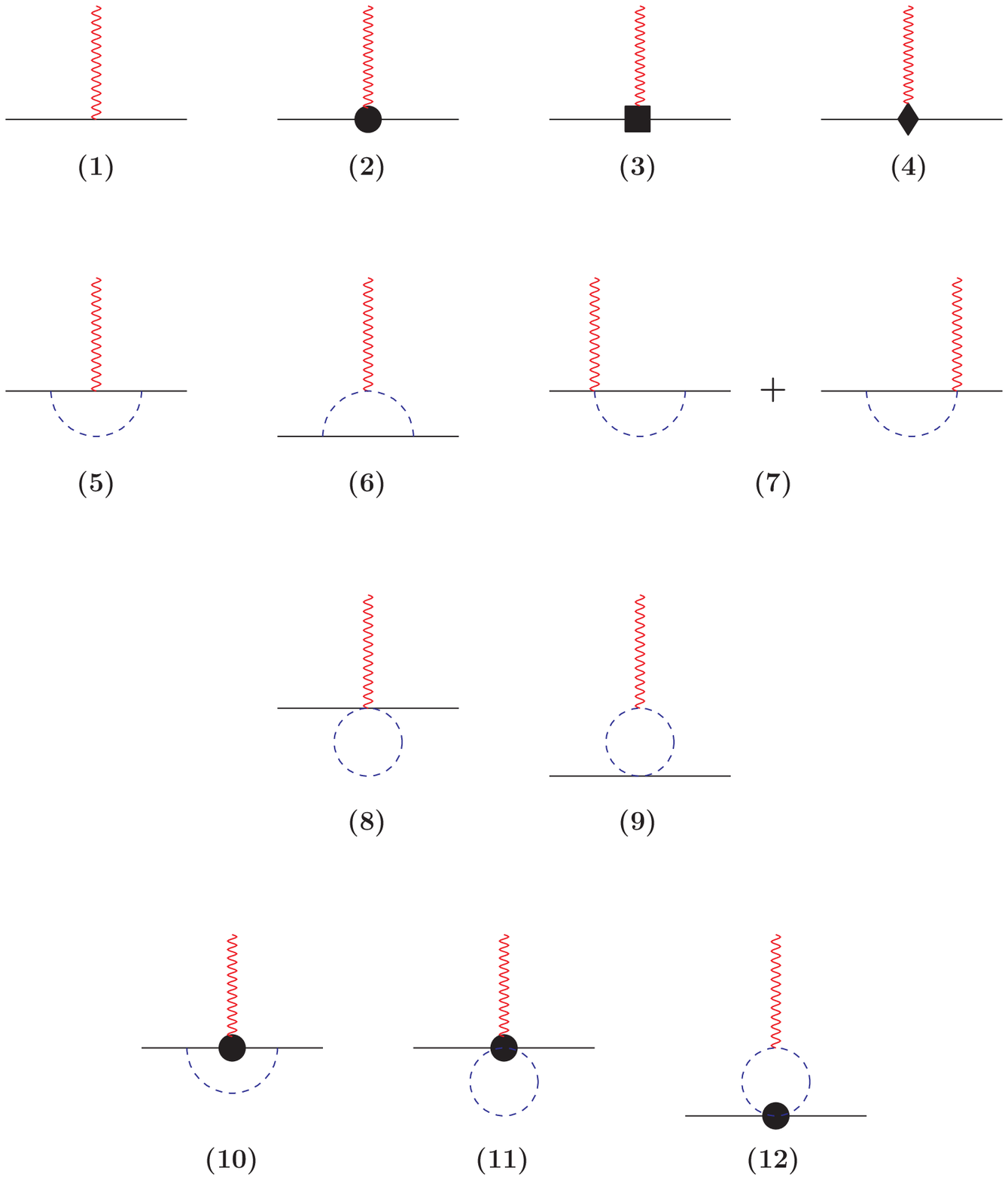,height=17cm}
\end{center}

\vspace*{0.5cm}

\noindent
{\bf Fig. 1.} {\em Diagrams including pseudoscalar meson contributions
to the EM quark transition operator up to fourth order.
Solid, dashed and wiggly lines refer to quarks, pseudoscalar mesons
and the electromagnetic field, respectively. Vertices denoted by a black
filled circle, box and diamond correspond to insertions from the second,
third and fourth order chiral Lagrangian.
\label{fig1}}

\newpage

\vspace*{.75cm}

\hspace*{2cm}
\begin{center}
\epsfig{file=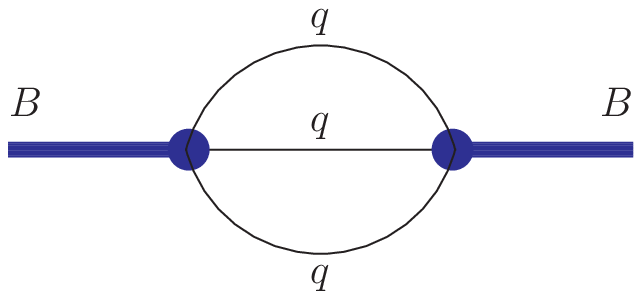,height=10cm}
\end{center}

\vspace*{-7cm}
\centerline{{\bf Fig.2} Baryon mass operator
\label{fig2}}

\vspace*{3cm}

\begin{center}
\epsfig{file=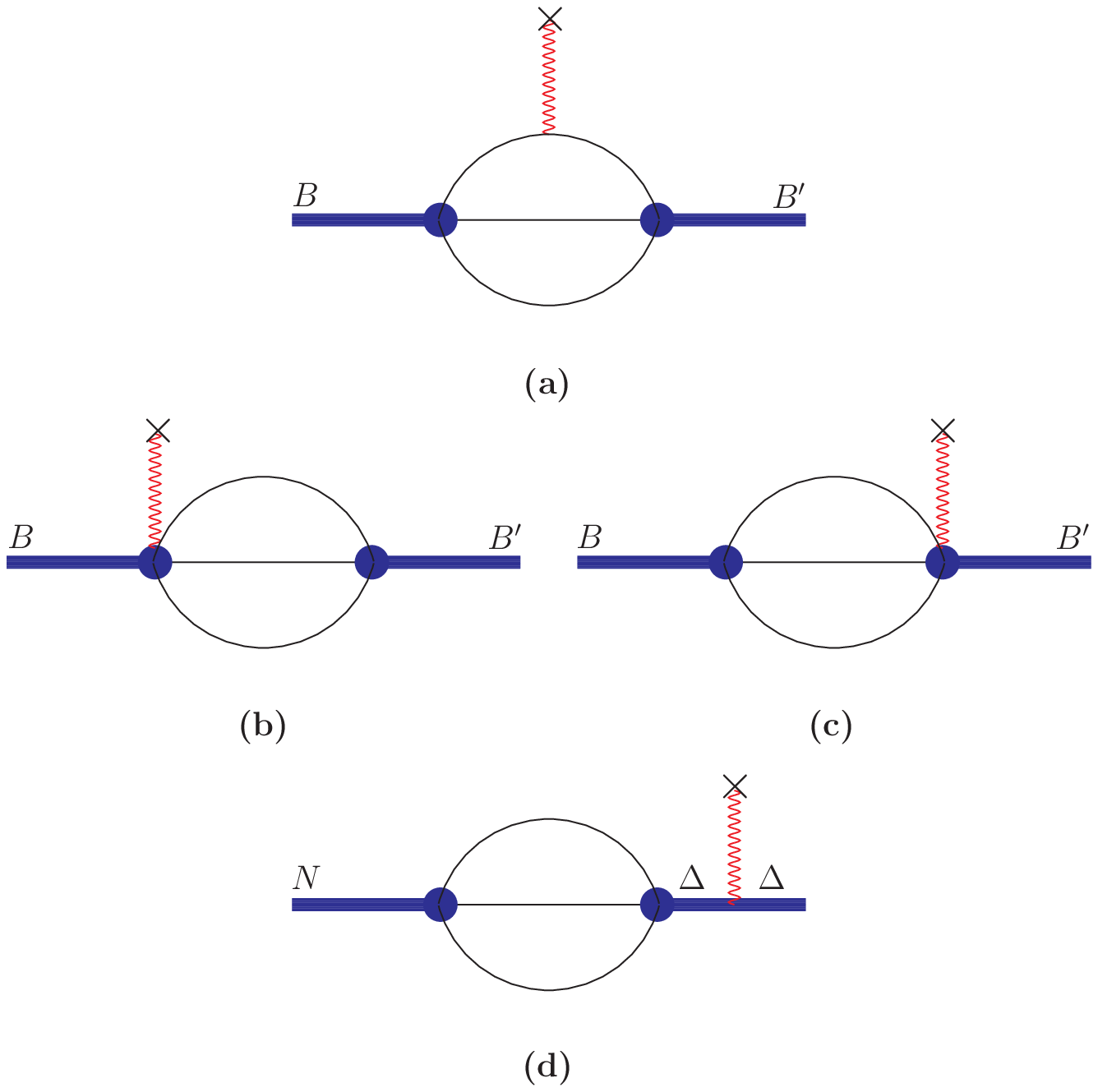,height=10cm}
\end{center}

\vspace*{.5cm}

\noindent {\bf Fig.3} {Diagrams contributing to the matrix elements
of the bare quark operators: triangle (a), bubble (b) and (c),
pole (d) diagrams. Symbol $\times$ corresponds to the source of
the external field.
\label{fig3}}

\newpage

\begin{center}
\epsfig{file=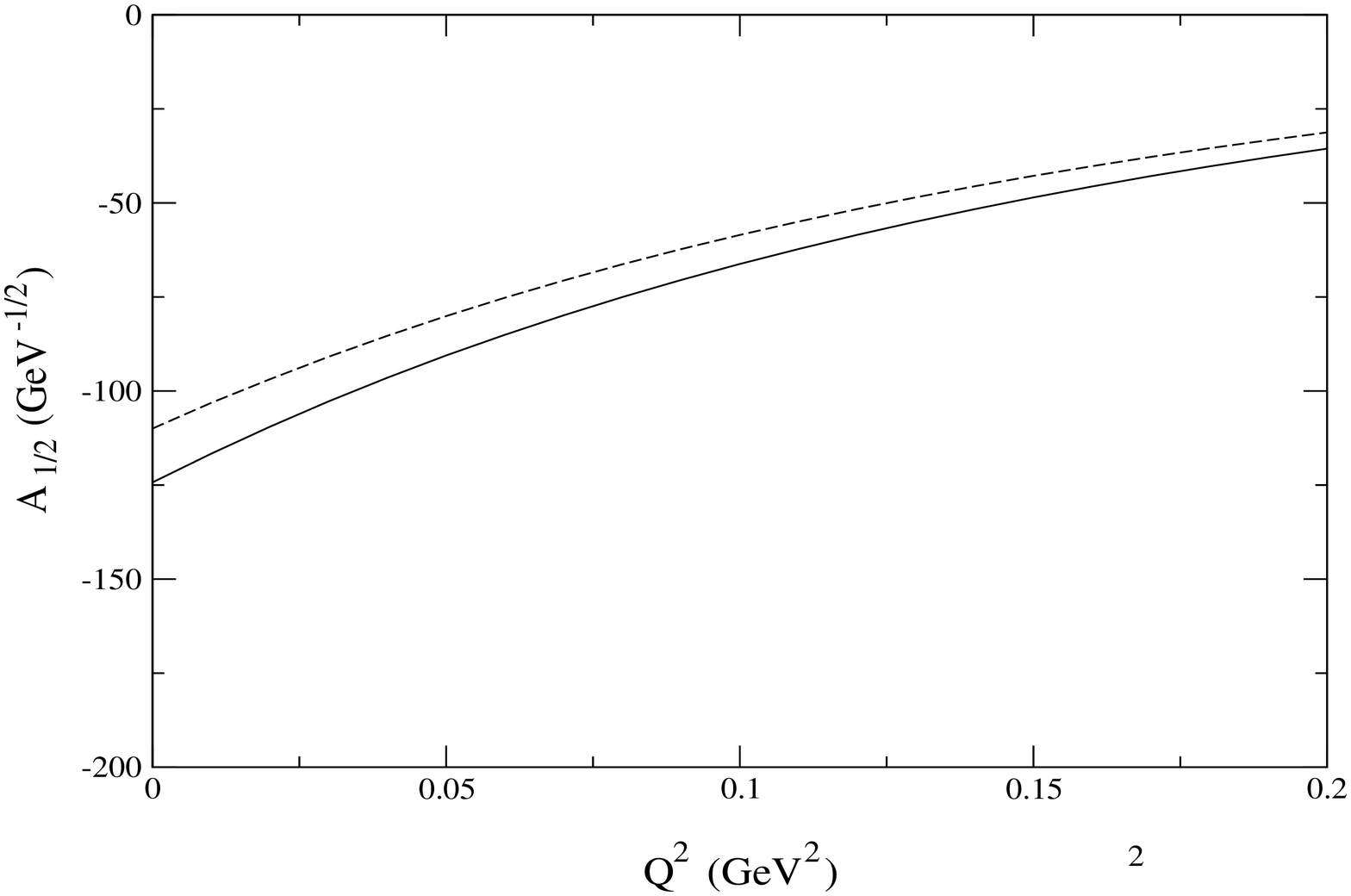,height=10cm}
\end{center}

\begin{center}
{\bf Fig.4}
Helicity amplitude $A_{1/2}(Q^2)$. Solid line is the total result, \\
whereas the dashed line
corresponds to the valence quark contribution
\label{fig4}
\end{center}

\begin{center}
\epsfig{file=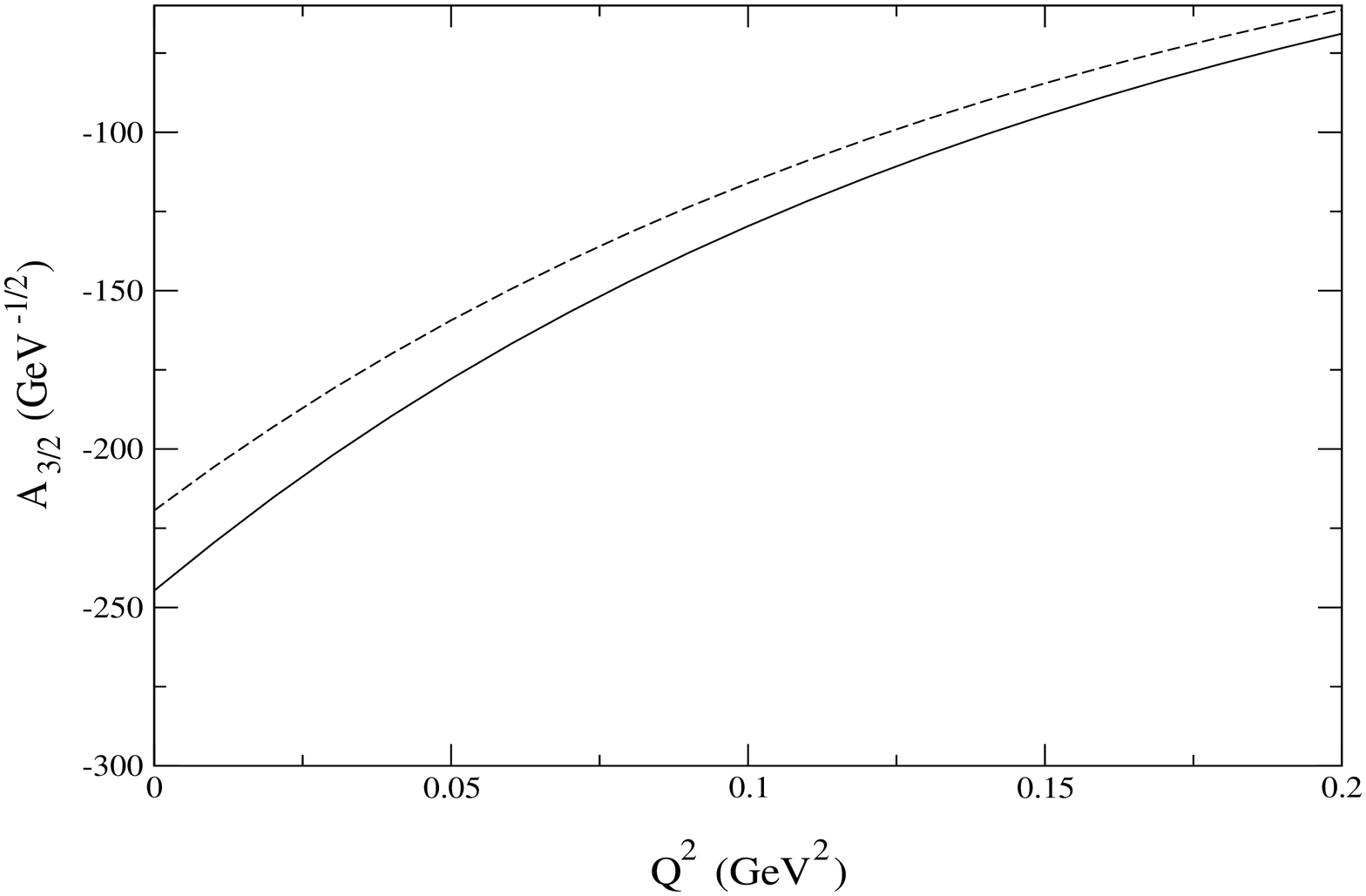,height=10cm}
\end{center}

\centerline{{\bf Fig.5} Helicity amplitude $A_{3/2}(Q^2)$.
Otherwise as in Fig.4.
\label{fig5}}

\newpage

\begin{center}
\epsfig{file=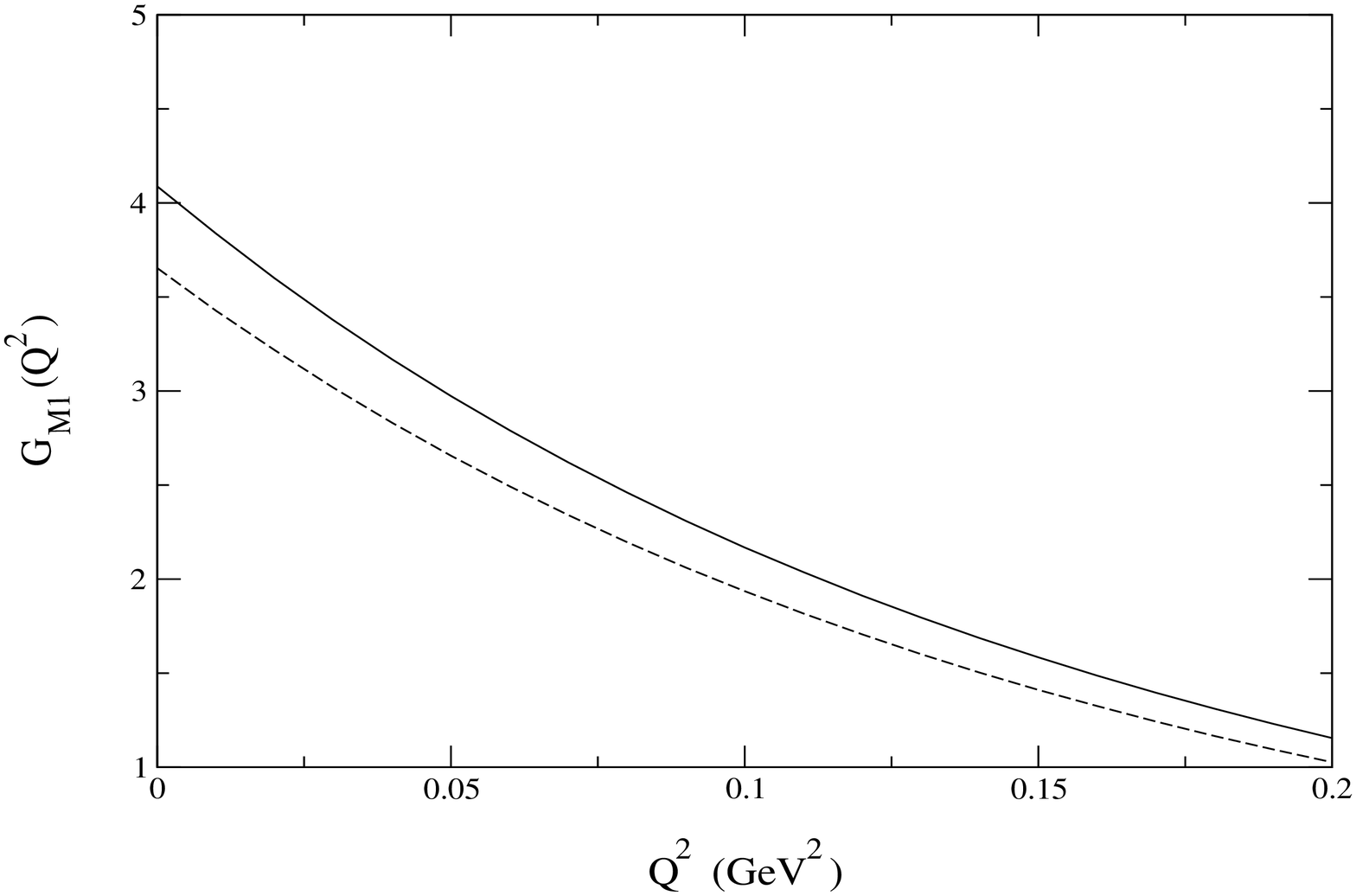,height=10cm}
\end{center}

\centerline{{\bf Fig.6} Form factor $G_{M1}(Q^2)$.
Otherwise as in Fig.4.
\label{fig6}}

\begin{center}
\epsfig{file=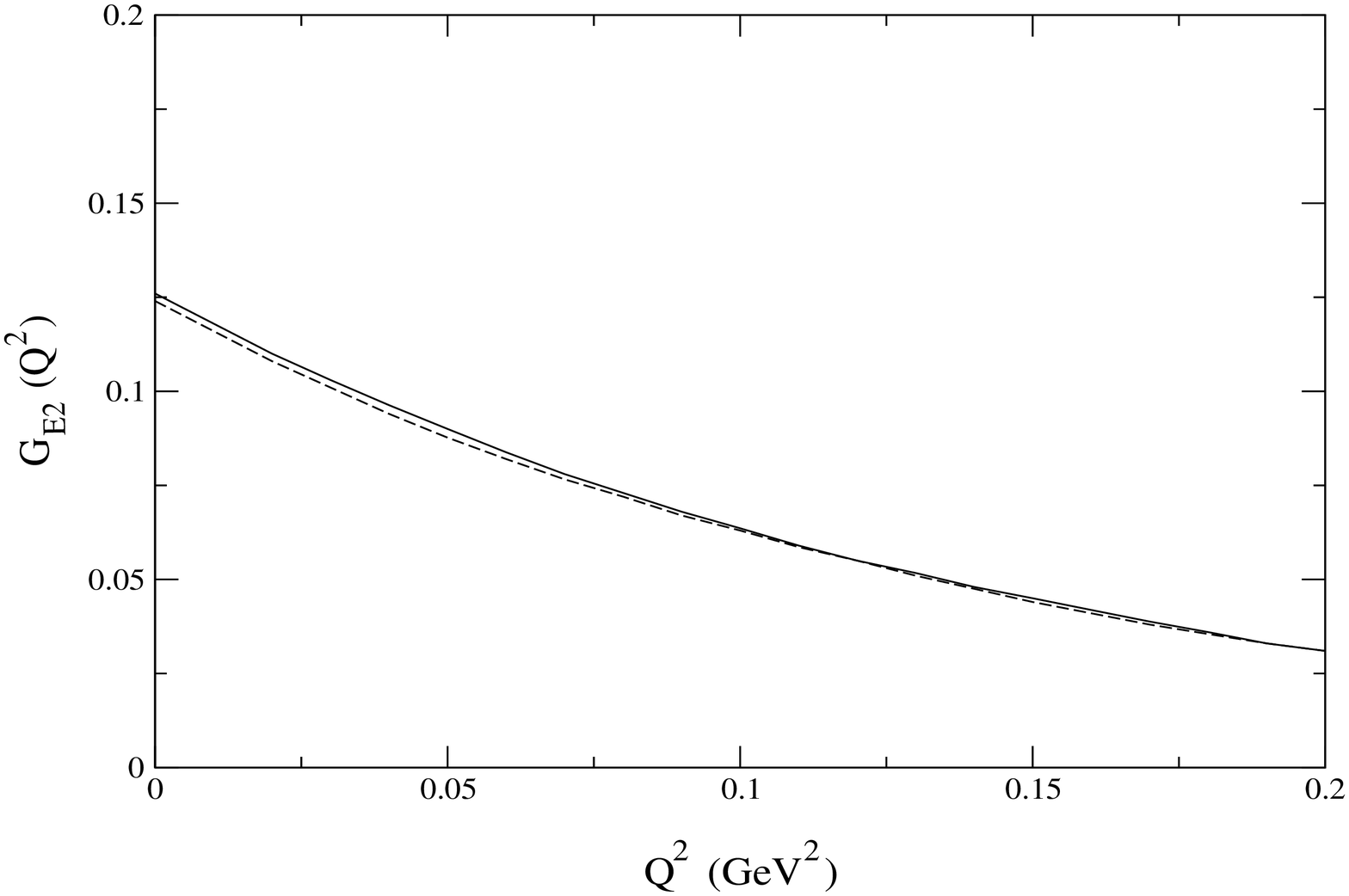,height=10cm}
\end{center}

\centerline{{\bf Fig.7} Form factor $G_{E2}(Q^2)$.
Otherwise as in Fig.4.
\label{fig7}}

\newpage

\begin{center}
\epsfig{file=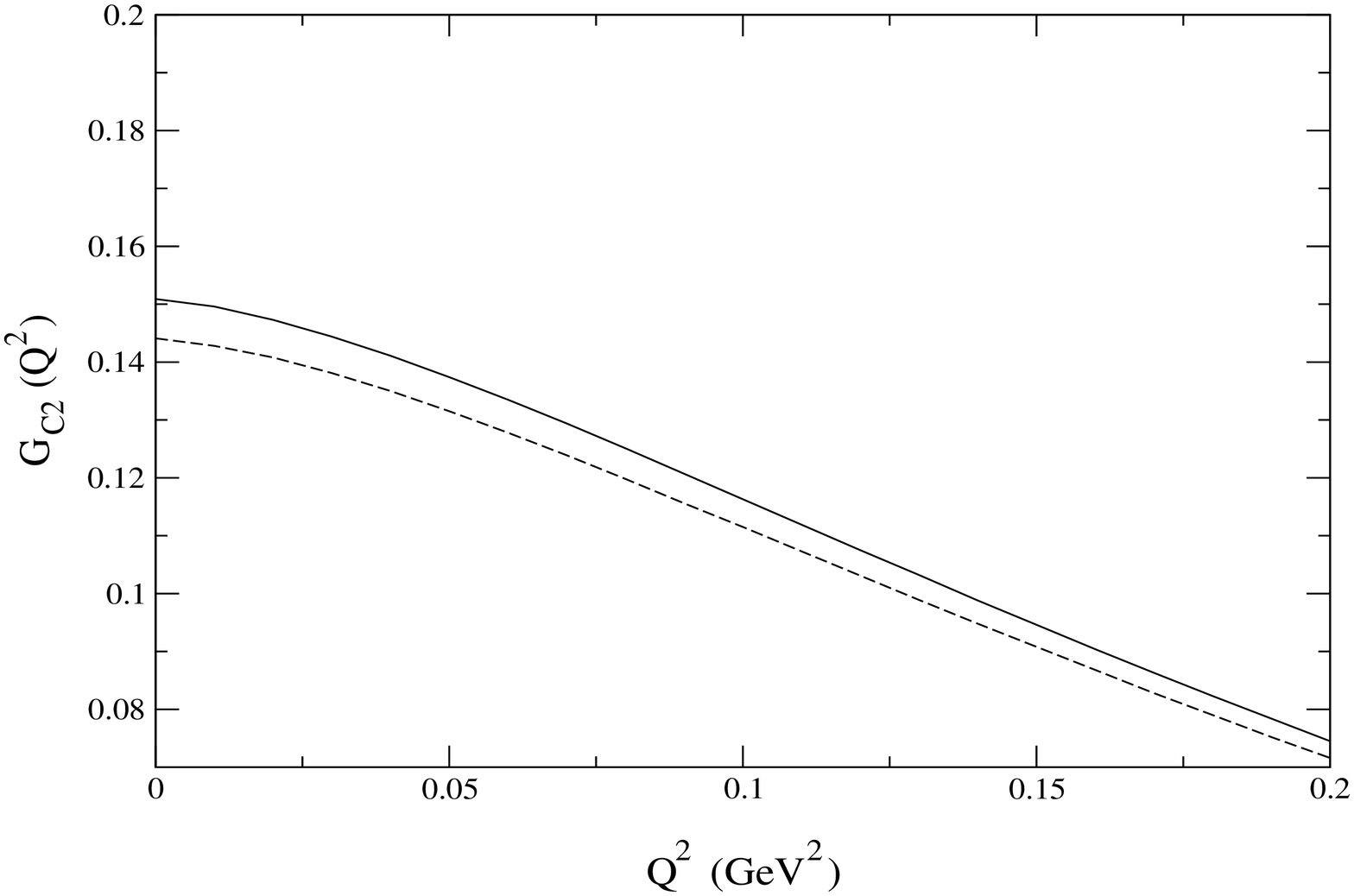, height=10cm}
\end{center}

\centerline{{\bf Fig.8} Form factor $G_{C2}(Q^2)$.
Otherwise as in Fig.4.
\label{fig8}}

\newpage

\begin{center}
\epsfig{file=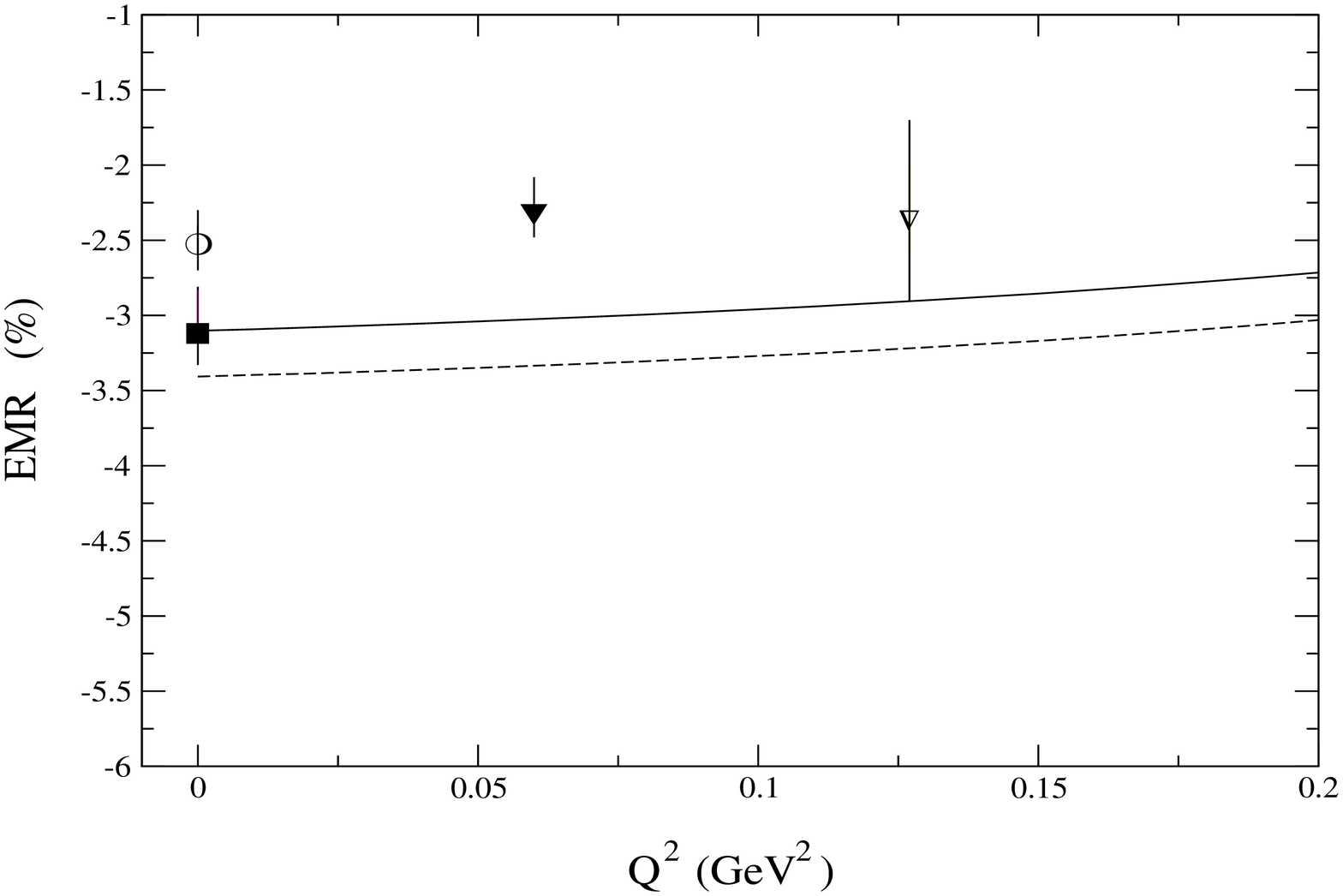,height=10cm}
\end{center}

{\bf Fig.9} Ratio EMR$(Q^2) = - G_{E2}(Q^2)/G_{M1}(Q^2)$.
Data are taken from Refs.~\cite{Stave:2006ea} (filled triangle),  \\
\hspace*{1.25cm} \cite{Blanpied:2001ae} (filled box),  
\cite{Beck:1999ge} (opened circle)  
and \cite{Sparveris:2004jn}
(opened triangle). Otherwise as in Fig.4.
\label{fig9}

\begin{center}
\epsfig{file=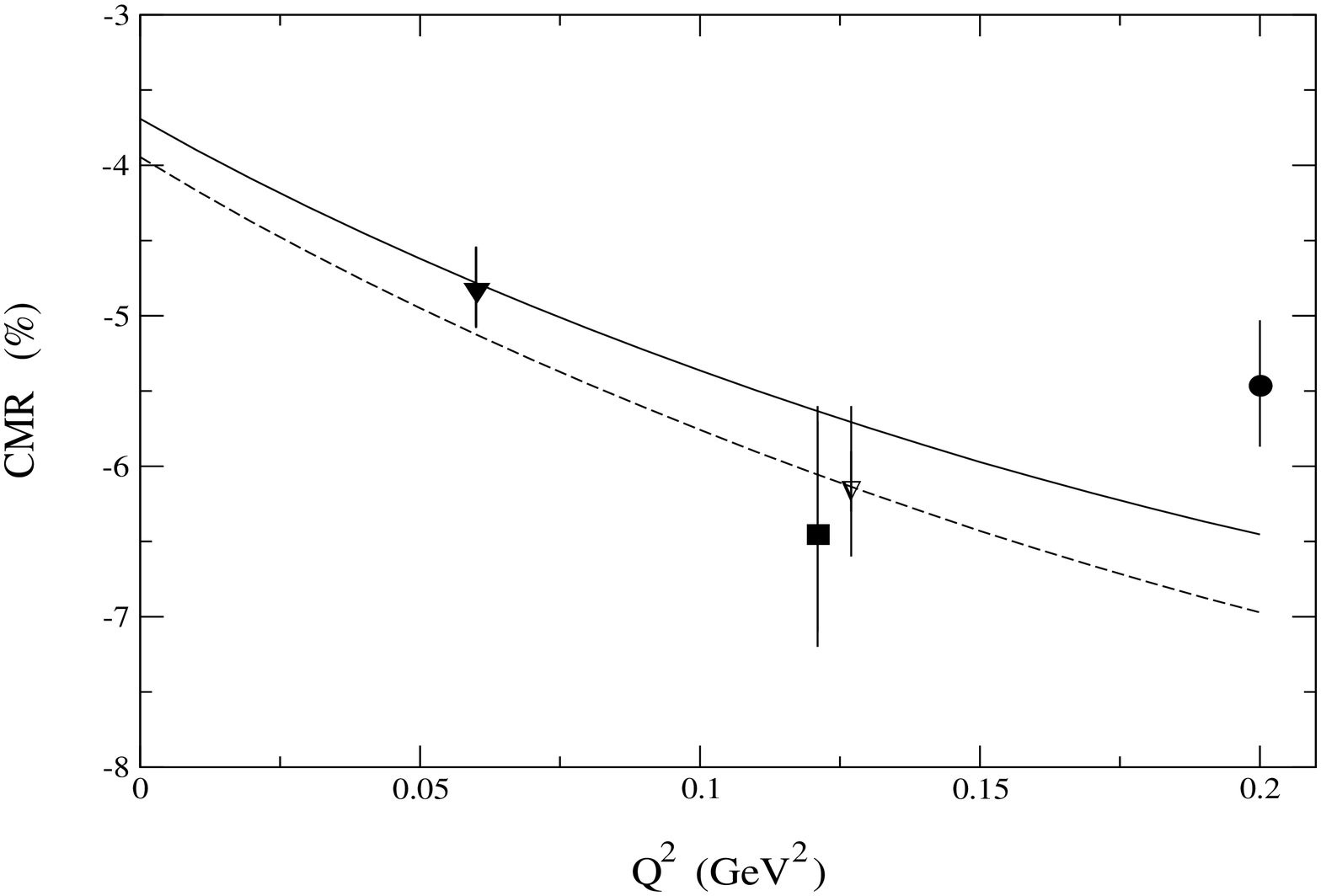,height=10cm}
\end{center}

{\bf Fig.10} Ratio CMR$(Q^2) = G_{C2}(Q^2)/G_{M1}(Q^2)$.
Data are taken from  Refs.~\cite{Stave:2006ea} (filled triangle), \\
\hspace*{1.25cm}  \cite{Sparveris:2004jn} (opened triangle),  
\cite{Pospischil:2000ad} (filled box) 
and \cite{Elsner:2005cz} (filled circle). Otherwise as in Fig.4.
\label{fig10}

\end{document}